\documentclass[useAMS,usenatbib]{mn2e}
\usepackage{graphicx,amssym}
\newcommand{\bc}{\begin{center}}
\newcommand{\ec}{\end{center}}

\title[The build--up of the colour--magnitude since $z \sim 0.8$]
      {The build--up of the colour--magnitude relation in galaxy clusters since
      $z\sim 0.8$\thanks{\small Based on observations collected at the European
      Southern Observatory, Chile, as part of large programme 166.A--0162 (the
      ESO Distant Cluster Survey)}} 
\author[G.~De Lucia et al.]
       {Gabriella De Lucia$^1$\thanks{Email: gdelucia@mpa-garching.mpg.de},
         Bianca M. Poggianti$^2$, Alfonso Arag\'on-Salamanca$^3$, 
	 \newauthor Simon D. M. White$^1$, Dennis Zaritsky$^4$, Douglas
	 Clowe$^4$, Claire Halliday$^5$, 
	 \newauthor Pascale Jablonka$^6$, Anja von der Linden$^1$, Bo
	 Milvang-Jensen$^7$, Roser Pell\'o$^8$, 
	 \newauthor Gregory Rudnick$^9$, Roberto P. Saglia$^{10}$, Luc
         Simard$^{11}$ 
         \\      
         $^1$Max--Planck--Institut f\"ur Astrophysik, 
         Karl--Schwarzschild--Str. 1, D-85748 Garching, Germany\\
         $^2$INAF-Osservatorio Astronomico di Padova, Vicolo dell'Osservatorio
	 5, Padova I-35122, Italy \\
	 $^3$School of Physics and Astronomy, University of Nottingham,
	 University Park, Nottingham NG7 2RD, UK \\
	 $^4$Steward Observatory, University of Arizona, 933 North Cherry
	 Avenue, Tucson, AZ 85721 \\
	 $^5$INAF,Osservatorio Astrofisico di Arcetri, Largo E. Fermi 5,
	 I-50125, Firenze, Italy \\  
	 $^6$Observatoire de Gen\`eve, Laboratoire d'Astrophysique Ecole 
	 Polytechnique Federale de Lausanne (EPFL), CH-1290 Sauverny,
	 Switzerland\\  
	 $^7$Dark Cosmology Centre, Niels Bohr Institute, University of
         Copenhagen, Juliane Maries Vej 30, DK-2100 Copenhagen, Denmark \\
	 $^8$Laboratoire d'Astrophysique, UMR 5572, Observatoire Midi-Pyrenees,
	 14 Avenue Edouard Belin, Toulouse F-31400, France \\
	 $^9$NOAO, 950 N. Cherry Ave, Tucson, AZ 85719, USA \\
         $^{10}$Max-Planck-Institut fur extraterrestrische Physik,
	 Giessenbachstrasse Postfach 1312, Garching D-85748, Germany\\
	 $^{11}$Herzberg Institute of Astrophysics, National Research Council
	 of Canada, 5071 West Saanich Road, Victoria, BC V9E 2E7, Canada 
       }
        
\begin{document}

%\date{Accepted 2006 ???? ??. 
%      Received 2006 ???? ??; 
%      in original form 2006 ???? ??}

\pagerange{\pageref{firstpage}--\pageref{lastpage}} 
\pubyear{2006}

\maketitle

\label{firstpage}

\begin{abstract}
  Using galaxy clusters from the ESO Distant Cluster Survey, we study how the
  distribution of galaxies along the colour--magnitude relation has evolved
  since $z\sim0.8$.  While red--sequence galaxies in all these clusters are
  well described by an old, passively evolving population, we confirm our
  previous finding of a significant evolution in their luminosity distribution
  as a function of redshift.  When compared to galaxy clusters in the local
  Universe, the high redshift EDisCS clusters exhibit a significant {\it
  deficit} of faint red galaxies.  Combining clusters in three different
  redshift bins, and defining as `faint' all galaxies in the range
  $0.4\,\gtrsim\,$L$/$L$_*\,\gtrsim\,0.1$, we find a clear decrease in the
  luminous--to--faint ratio of red galaxies from $z\sim 0.8$ to $z\sim 0.4$.
  The amount of such a decrease appears to be in qualitative agreement with
  predictions of a model where the blue bright galaxies that populate the
  colour--magnitude diagram of high redshift clusters, have their star
  formation suppressed by the hostile cluster environment. Although model
  results need to be interpreted with caution, our findings clearly indicate
  that the red--sequence population of high--redshift clusters does not contain
  all progenitors of nearby red--sequence cluster galaxies.  A significant
  fraction of these must have moved onto the red--sequence below $z\sim 0.8$.
\end{abstract}


%%%%%%%%%%%%%%%%%%%%%%%%%%%%%%%%%%%%%%%%%%%%%%%%%%%%%%%%%%%%%%%%%%%%%%%%%%%%%%%
\section{Introduction}
\label{sec:intro}

Galaxy clusters may be considered as laboratories for studying the physical
processes that drive galaxy evolution.  They offer the possibility to trace the
properties of galaxies in similar environments over a relatively long time
baseline.  In addition, they offer the practical advantage of providing many
galaxies in a relatively small region of the sky and all approximately at the
same redshift. This allows efficient observation even with modest fields of
view and modest amounts of telescope time.  It should be noted, however, that
in order to establish that physical processes related to the cluster
environment are indeed playing a role, it is necessary to compare the evolution
of similar galaxies in different environments (i.e. in the clusters and in the
`field').  In addition, galaxy clusters represent a {\it biased} environment
for evolutionary studies.  In the current standard cosmogony, clusters
originate from the gravitational collapse of the highest (and rarest) peaks of
primordial density perturbations, and evolutionary processes in these regions
occur at an accelerated pace with respect to regions of the Universe with {\it
average} density.

The technical capabilities achieved in recent years have provided a rapidly
growing database on high--redshift clusters
\citep{zaritsky97,gonzalez01,valtchanov04,gladders05,kodama05,white05,wilson06}.
These, interpreted using the latest theoretical techniques
\citep{cole00,hatton03,springel05,delucia06}, should provide important
constraints on the physical mechanisms driving the formation and the evolution
of cluster galaxies.

Early studies of the galaxy population in distant clusters pointed out
significant differences with respect to nearby systems \citep{bo84}.  More
recent work has provided us with a much more detailed picture of these
differences \citep[a very incomplete list of recent papers
includes][]{stanford05,postman05,jorgensen05,white05,poggianti06,strazzullo06}.
One interesting outcome of these studies has been the discovery that a tight
relation between the colours and the magnitudes of bright elliptical galaxies
holds up to the highest redshifts probed so far
\citep{blakeslee03,delucia04b,holden04,mei06}.

The existence of a colour--magnitude relation (hereinafter CMR) has been known
for a long time \citep{dV61,vs77}.  At least in nearby clusters
\citep{depropris98}, it appears to extend $5 - 6$ mag faint-ward of the
Brightest Cluster Galaxy (BCG).  At the present epoch, the CMR can be
interpreted either as a result of differing {\it ages} (bluer galaxies being
younger), or of differing {\it metallicities} (bluer galaxies being more metal
poor), or as a combination of the two \citep[Ferreras, Charlot \& Silk 1999;][
and references therein]{terlevich99,poggianti01a}.  

The mere existence of a tight relation at high redshift favours the metallicity
interpretation and naively appears to make the age explanation untenable.  The
reason for this is that if one {\it assumes} that all present--day
red--sequence galaxies are still identified as red--sequence members in
high--redshift clusters, then if the CMR were primarily age driven it would
change dramatically with increasing redshift as small galaxies approach their
formation epoch and progressively become brighter and bluer \citep{kodama98}.
This expectation is in contrast with observational results which show that the
slope of the CMR does not change appreciably over the redshift interval $0 - 1$
\citep[][and more recent work mentioned above]{gladders98,stanford98}.

One possible simple interpretation is then that cluster elliptical galaxies
represent a passively evolving population formed at high-redshift (z $> 2-3$)
in a short duration event \citep*[but see the discussion by][]{bower98}.  In
this scenario -- often referred to as {\it monolithic} -- the CMR arises
through the effects of supernovae winds: supernovae explosions heat the
interstellar medium triggering galactic winds whenever the thermal energy of
the gas exceeds its gravitational binding energy.  Since smaller galaxies have
shallower potential wells, this results in greater mass loss by smaller
systems, naturally establishing the observed CMR.  A difficulty may be that the
observed CMR shows no sign of a turnover at high mass of the kind predicted by
such models \citep{larson74}.

This simple model may be too naive for explaining the origin and the evolution
of the observed CMR.  In the {\it monolithic scenario} a galaxy has a single
well--defined progenitor at each redshift and its evolution is described by
simple smooth functions of time.  This is not true in the current standard
cosmological paradigm, where a single galaxy today corresponds to the ensemble
of all its progenitors at any previous redshift \citep[see discussion in
][]{deluciablaizot06}.  It is not obvious then that high--redshift red-sequence
galaxies contain all or even most of the progenitors of nearby red--sequence
cluster galaxies.  Indeed the results we present below give direct evidence
that this is not the case.  In addition, the simple model described above,
clearly neglects the infall of ``new'' galaxies during cosmological growth of
the cluster.  If the intra-cluster environment is associated with suppression
of star formation, then these galaxies would become redder and fainter and
might also join the red--sequence galaxy population at lower redshifts.

An alternative scenario has been proposed by \citet{kc98} - see also
\citet*{delucia04a} - in the framework of hierarchical models of galaxy
formation.  In these models, the CMR arises as a result of the fact that more
massive ellipticals originate from the mergers of more massive - and more
metal--rich - disk systems.  The models show a well defined red--sequence,
still mainly driven by metallicity differences, that is in place up to redshift
$\sim 2$, although with a scatter that is larger than that observed.

Recent work on the observed colour--magnitude relation of high--redshift
clusters has pointed out a new and still controversial result concerning an
apparent `truncation' of the CMR at redshift about $0.8$
\citep{delucia04b,kodama04}.

In \citet{delucia04b}, we analysed the colour--magnitude relation of $4$
clusters in the redshift interval $0.7$--$0.8$ from the ESO Distant Cluster
Survey (hereinafter EDisCS) and found a deficiency of low luminosity passive
red galaxies with respect to the nearby Coma cluster.  A decrease in the number
of faint red galaxies was detected in all clusters under investigation but one
(with low number of cluster members), although the significance of the deficit
was only at about the $3\sigma$ level.  In this paper we extend our analysis to
the full EDisCS sample covering the redshift range $0.4 - 0.8$ and a wide range
of structural properties.  The plan of the paper is as follows.  The
observational data used for our study are briefly described in
Sec.~\ref{sec:observations}.  In Sec.~\ref{sec:members} we present the criteria
used to define cluster membership and in Sec.~\ref{sec:cm} we present the
colour--magnitude relation for all the clusters in the EDisCS sample.  In
Sec.~\ref{sec:redshift} we study the distribution of galaxies along the
red--sequence, and discuss its dependence on redshift and on cluster velocity
dispersion.  The red--sequence galaxy distribution in nearby clusters is
studied in Sec.~\ref{sec:local}. In Sec.~\ref{sec:build}, we interpret the
evolution measured as a function of redshift in terms of simple population
synthesis models.  Finally, in Sec.~\ref{sec:discconcl}, we discuss our results
and give our conclusions.

Throughout this paper we will assume a $\Lambda$CDM cosmology: $H_0 = 70\,{\rm
km}\,{\rm s}^{-1}\,{\rm Mpc}^{-1}$, $\Omega_{\rm m} = 0.3$ and
$\Omega_{\Lambda}=0.7$. With this cosmology, $z \sim 0.8$ - the highest
redshift probed by our cluster sample - corresponds to more that $50$ per cent
of the look-back time to the Big Bang. Throughout this paper we use Vega
magnitudes, unless otherwise stated.
%%%%%%%%%%%%%%%%%%%%%%%%%%%%%%%%%%%%%%%%%%%%%%%%%%%%%%%%%%%%%%%%%%%%%%%%%%%%%%%
\section{The data}
\label{sec:observations}

EDisCS is an ESO Large Programme aimed at the study of cluster structure and
cluster galaxy evolution over a significant fraction of cosmic time.  The
complete EDisCS dataset provides homogeneous photometry and spectroscopy for
$20$ fields containing galaxy clusters at $z=0.4$--$1$.  Clusters candidates
were selected from the Las Campanas Distant Cluster Survey (LCDCS) of
\citet{gonzalez01} by identifying surface brightness excesses using a very wide
filter ($4500$--$7500$ \AA) in order to maximise the signal-to-noise of distant
clusters against the sky.  The EDisCS sample of $20$ clusters was constructed
selecting $30$ from the highest surface brightness candidates in the LCDCS, and
confirming the presence of an apparent cluster and of a possible red sequence
with VLT $20$min exposures in two filters.  From these $30$ candidates, we then
followed up $10$ among the highest surface brightness clusters in the LCDCS in
each of the ranges of estimated redshift $0.45 < z_{\rm est} < 0.55$ and $0.75
< z_{\rm est} < 0.85$. In the following, we will often refer to these as the
intermediate and high redshift samples respectively.

As a consequence of the scatter of the estimated redshifts about the true
value, we have ended up with a set of clusters distributed relatively smoothly
between $z=0.42$ and $z=0.96$, rather than two samples concentrated at $0.5$
and $0.8$, as originally planned.  Details on the selection of cluster
candidates can be found in \citet{white05}.  Our follow-up programme obtained
deep optical photometry with FORS2/VLT \citep{white05}, near-IR photometry with
SOFI/NTT (Arag{\'o}n-Salamanca et al. in preparation), and multislit
spectroscopy with FORS2/VLT for the $20$ fields (Halliday et al. 2004;
Milvang-Jensen et al. in preparation).  ACS/HST mosaic imaging of $10$ of the
highest redshift clusters has also been acquired (Desai et al. in preparation).
For three EDisCS clusters, narrowband $H_{\alpha}$ imaging has been acquired
\citep{finn05} and for three clusters we have XMM data \citep{Johnson06}.

The optical ground--based photometry and a first basic characterisation of our
sample of clusters as a whole, is presented in \citet{white05}.  In brief, our
optical photometry consists of V, R, and I imaging for the $10$ highest
redshift cluster candidates and B, V, and I imaging for the remaining $10$
intermediate redshift cluster candidates.  Total integration times were
typically $45$ minutes at the lower redshift and $2$ hours at the higher
redshift.  Object catalogues have been created using the SExtractor software
version 2.2.2 \citep{ba96} in `two--image' mode using the I--band images as
detection reference images.  Magnitudes and colours have been measured on the
seeing--matched images (to $0\farcs8$ -- the typical seeing in our IR images)
using fixed circular apertures.  Throughout this paper, we correct magnitudes
and colours for Galactic extinction according to the maps of
\citet*{schlegel98} and a standard Milky Way reddening curve. We refer to
\citet{white05} for details about our photometry.  In the following, we will
use magnitudes and colours measured using a fixed circular aperture with
$1\farcs0$ radius.  This choice has been adopted to simplify the comparison
with the Coma cluster, as we will explain later in the paper.  The cluster
velocity dispersions and ${\rm R}_{200}$ we use in the following are the same
as used in \citet{poggianti06} and are listed in Table~1 of that paper.

As a part of our programme, we also obtained spectra for $> 100$ galaxies per
cluster field (typical exposure times were $4$ hours for the high redshift
candidates and $2$ hours for the intermediate redshift clusters).  The
spectroscopic selection, observations, data reduction, and spectroscopic
catalogues are presented in \citet{halliday04} and Milvang-Jensen et al. (in
preparation).  As explained in \citet{white05}, deep spectroscopy was not
obtained for two of the EDisCS fields (cl$1122.9$-$1136$\footnote{Only one
  short exposure mask was obtained for this field, showing no evidence of a
  concentration of galaxies at any specific redshift.} and
cl$1238.5$-$1144$\footnote{Only two short exposure masks were obtained for this
  field for which we also do not have NIR data.}), which are not included in
the present study.

%%%%%%%%%%%%%%%%%%%%%%%%%%%%%%%%%%%%%%%%%%%%%%%%%%%%%%%%%%%%%%%%%%%%%%%%%%%%%%%

\section{Cluster membership}
\label{sec:members}

Although complications arise from the existence of redshift space distortions,
spectroscopic redshifts provide the optimal technique to determine cluster
membership. However, obtaining spectroscopic redshifts for large numbers of
faint objects is not feasible within the available time with current
instrumentation, even for samples just beyond $z = 0.5$.  A standard method to
correct for field contamination, in absence of spectroscopy, is to use
statistical field subtraction \citep{as93,stanford98,kb01}: a `cluster--free'
field is used to determine the number of contaminating galaxies as a function
of magnitude and/or colour.  This method becomes increasingly uncertain at high
redshift: \citet{driver98}, for example, used simulations to show that it is
already unreliable at $z > 0.3$.  In addition, this approach does not provide
the likelihood of being a cluster member on a galaxy--by--galaxy basis.

In the last decade, the techniques used to determine photometric redshifts have
become much more precise, suggesting that they can be used to address specific
scientific questions \citep{benitez00,bolzonella00,rudnick01,firth03}.  An
important by--product is an estimate of the spectral type for each observed
galaxy.  Errors in estimated photometric redshifts are much larger than typical
errors in spectroscopic redshifts.  In addition, systematics or degeneracies
are often present because of uncertainties in the redshift evolution of
spectral energy distributions and/or insufficient calibrating spectroscopy for
the magnitude range sampled by the photometric data.

Given the difficulties mentioned above, we decided in the present study to use
both a `classical' statistical field subtraction and a membership criterion
based on photometric redshift information.  We give a brief description of both
methods in the following sections.

\subsection[]{Photometric redshifts}
\label{sec:photoz}

Photometric redshifts were computed using two different codes
\citep{bolzonella00,rudnick01} in order to provide better control of the
systematics in the identification of likely non--members.  The two codes
employed in this study are based on the use of similar SED fitting procedures,
but different template spectra, different minimisation algorithms, and a number
of other different details (see the original papers).  The performance of these
codes on the EDisCS dataset will be examined in Pell\'o et al. (in
preparation).

For the purposes of this analysis, the codes were run allowing a maximum
photometric redshift of $2$ and assuming a $5$ per cent minimum flux error for
the photometry.  Where they can be checked, the photometric redshifts of the
galaxies in our sample are quite accurate with $<|z_{\rm spec} - z_{\rm phot}|>
= 0.06$--$0.08$.  There is no systematic trend between $z_{\rm phot}-z_{\rm
spec}$ and $z_{\rm spec}$ and the percentage of catastrophic failures, i.e. the
fraction of objects with $|z_{\rm spec} - z_{\rm phot}| > 0.3$, is of the order
of $10$ per cent.

In the present study, we use the redshift probability distributions provided by
the two photometric redshift codes, as a quantitative tool to estimate cluster
membership.  Briefly, we accept galaxies as potential cluster members if the
integrated probability for the photometric redshift to be within $\pm\,0.1$ of
the (known) cluster redshift is greater than a specific threshold for {\it
both} of the photometric redshift codes.  These probability thresholds (${\rm
P}_{\rm thresh}$) range from $0.1$ to $\,0.35$, depending on the filter set
available for each particular field, and they were calibrated using our
spectroscopy to maximise the cluster membership and, at the same time, to
minimise contamination from interlopers.

Calibration against our spectroscopic sample shows that this technique allows
us to retain more than $90$ per cent of the cluster members while rejecting
slightly less than $50$ per cent of the non--members in the spectroscopic
sample.  The efficiency of rejection for the bluest and reddest halves of the
sample is similar to within less than $10$ per cent.  However, red non--members
are slightly more efficiently rejected, and red members slightly more
efficiently accepted than their blue counterparts at identical ${\rm P}_{\rm
  thresh}$.  This is expected because of the poorer constraints on $z_{\rm
  phot}$ for blue galaxies due to their smoother SEDs.

A method similar to that used in our study was proposed by \citet{bl00}. Their
application was based on the use of an ``empirical'' photometric redshift
technique.  The latter is based on the use of an empirical relation, measured
for the spectroscopic sample available, between the spectroscopic redshifts and
the photometric data points. The use of this method, forced Brunner \& Lubin to
\emph{assume} a probability distribution, which they supposed to be Gaussian
with mean given by the estimated photometric redshift and standard deviation
defined by the estimated error.  The error distributions are, however, usually
strongly \emph{non--Gaussian} \citep{rudnick01}, so proper use of the
probability distributions provides a better estimate of the real uncertainty in
the photometric redshift estimates.

When available, we use the spectroscopic information to determine cluster
membership: spectroscopic non-members that are erroneously classified as
cluster members by the photometric redshift criterion detailed above, are
rejected.  Spectroscopic members that are erroneously rejected are re-included
into the sample before the analysis.  The photometric redshift technique we use
performs well, particularly on red galaxies.  As a consequence, this latter
correction based on the availability of spectroscopic information, does not
modify significantly the results discussed below.  We note that spectroscopic
membership has been defined as in \citet{halliday04}.  The number of
spectroscopic members for our clusters ranges from $11$ to $67$ \citep[see
Table~1 in][]{poggianti06}.  If membership is assigned using the photometric
redshift method outlined above, the fraction of cluster members for which
spectra are available ranges from a few to about $15$ per cent.

\subsection[]{Statistical subtraction}
\label{sec:statsub}

As an alternative method to determine cluster membership, we employ a
`classical' statistical subtraction technique.  The method we use is similar to
that adopted in \citet{pimbblet02}.  We refer to the original paper for more
details on the procedure which we only briefly outline here.

The `field' population has been determined from one field of the Canada France
Deep Field Survey \citep{mccracken01}\footnote{The catalogue has been kindly
  provided to us by H. McCracken}.  This corresponds to an area of about
$0.25\,$deg$^2$, which is much larger than the cluster area used for our
analysis (see next section).  Both the cluster and the field regions are binned
onto a gridded colour--magnitude diagram (we use a $0.3$ bin in colour and a
$0.5$ bin in magnitude).  The field region is then scaled to the same area as
the cluster region we wish to correct, and each galaxy is assigned a
probability to be a cluster member simply by counting how many galaxies lie in
the colour--magnitude bin in the two different regions.  Using a Monte Carlo
method, the field population is then subtracted off.  If this procedure gives a
negative number of galaxies in the cluster population at a particular grid
position, the mesh size is increased for that particular position.  As
explained in Appendix A of \citet{pimbblet02}, this approach has the advantage
of preserving the original probability distribution better than similar methods
\citep{kb01} where the excess probability is distributed evenly between the
neighbours of critical grid positions.  For each cluster we run $100$ Monte
Carlo realizations of the above procedure.

When available, we use the spectroscopic information: spectroscopic members and
non-members are always assigned a probability $1$ and $0$ to be cluster members
respectively.  As is the case when cluster membership is assigned using
photometric redshifts, this correction does not modify significantly the
results presented below.

%%%%%%%%%%%%%%%%%%%%%%%%%%%%%%%%%%%%%%%%%%%%%%%%%%%%%%%%%%%%%%%%%%%%%%%%%%%%%%%
\section{The colour--magnitude relation}
\label{sec:cm}

\begin{figure*}
\bc
\resizebox{16cm}{!}{\includegraphics{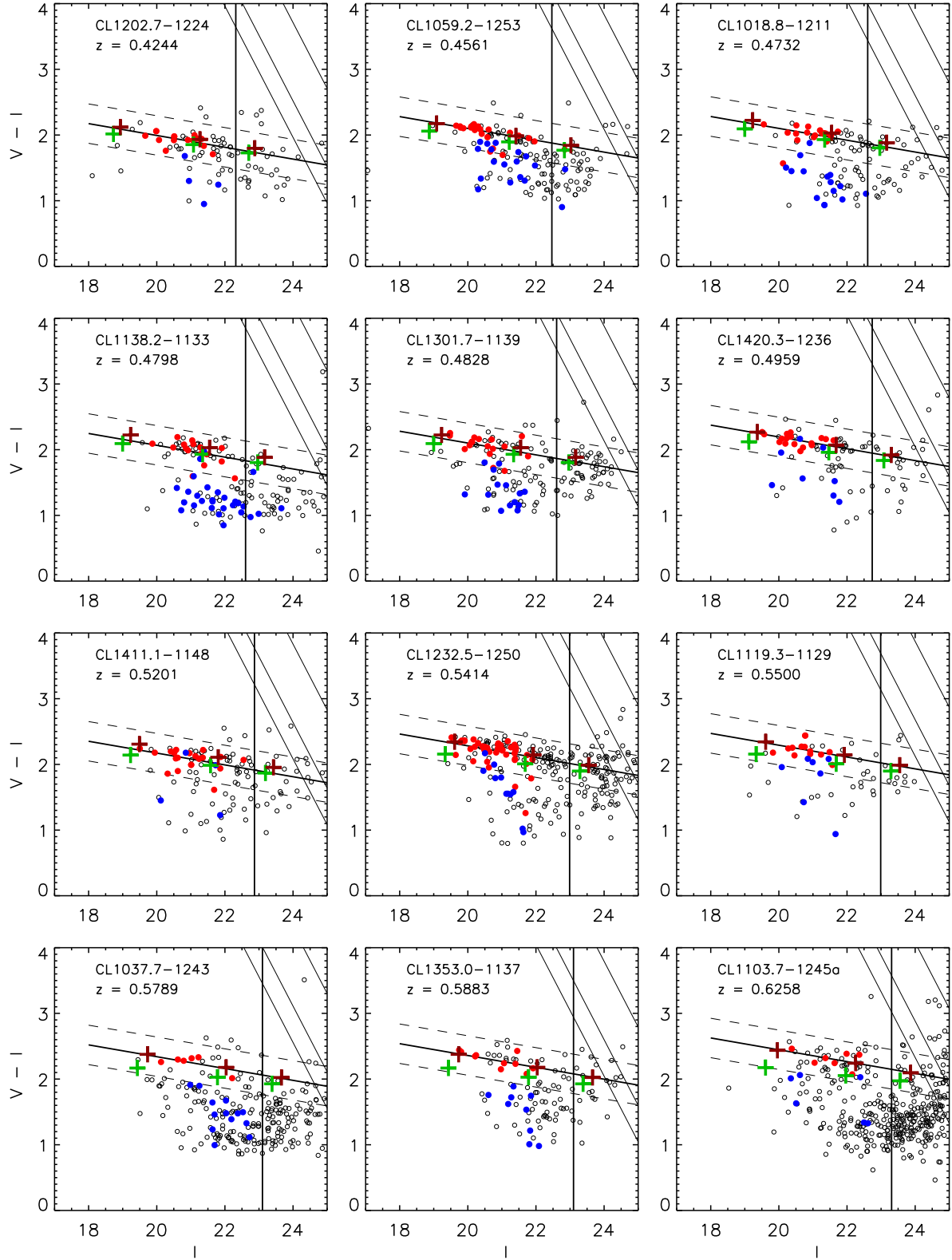}}\\%
\caption{Colour--magnitude diagrams for the $18$ EDisCS fields used in this
  study.  Empty circles show objects retained by our photometric redshift
  criterion.  Blue and red filled circles represent spectroscopically confirmed
  members with and without emission lines in their spectra.  Thin slanted lines
  correspond to the $1$, $3$ and $5\sigma$ detection limits in the V-band.  The
  solid thick line in each panel, represents the best fit relation measured
  using the bi--weight estimator and assuming a fixed slope of $-0.09$.  Dashed
  lines correspond to $\pm\,0.3$~mag from the best fit line. Crosses show the
  location of galaxy models with two different SF histories.  The solid
  vertical line in each panel shows the apparent magnitude which translates to
  M$_{\rm V}=-18.2$ when evolved passively to $z=0$. (See text for details).  }
\label{fig:cma}
\ec
\end{figure*}
\addtocounter{figure}{-1}
\begin{figure*}
\bc
\resizebox{16cm}{!}{\includegraphics{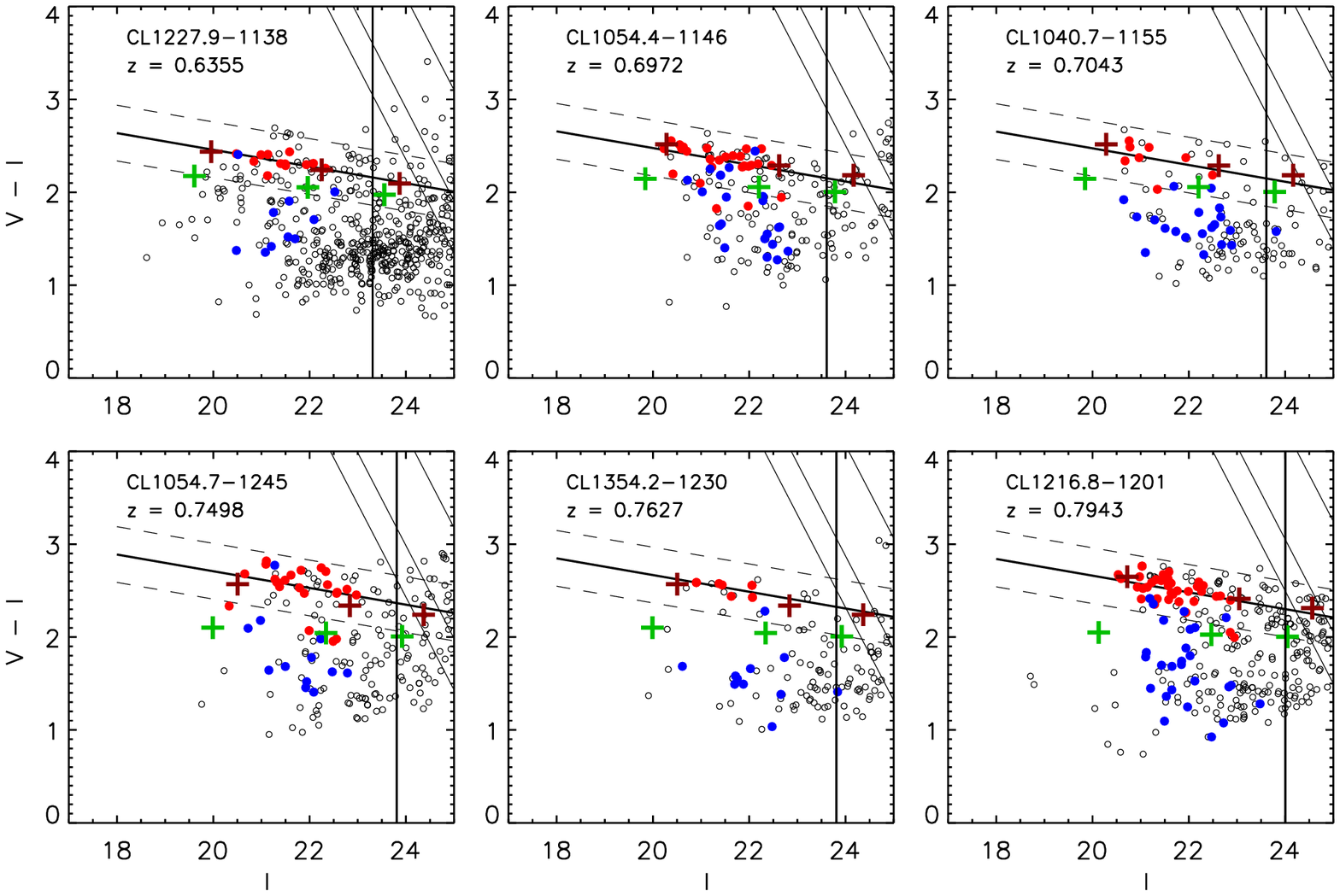}}\\%
\caption{{\bf -- continued}}
\label{fig:cmb}
\ec
\end{figure*}

In Fig.~\ref{fig:cma} we show the colour-magnitude diagrams for the $18$ fields
for which we have high quality spectroscopy.  Clusters are shown in order of
increasing redshift. Empty circles show objects for which our photometric
redshift criterion gives a high probability of cluster membership (see
Sec.~\ref{sec:photoz}).  Red and blue filled circles represent
spectroscopically confirmed members lacking/showing any emission line in their
spectra respectively.  We note that the typical detection limit for the
[OII]3727 line is low in our spectra - approximately
$2\,$\AA~\citep{poggianti06}.  Thin slanted lines correspond to the $1$, $3$
and $5\sigma$ detection limits in the V-band, while the solid thick line in
each panel represents the best fit to the colour--magnitude relation measured
using a fixed slope of $-0.09$.  The fit has been computed using the bi--weight
estimator \citep*{beers00} on the objects without emission lines in their
spectra (red filled circles).  Dashed lines correspond to $\pm\,0.3$~mag from
the best fit line.

Crosses in Fig.~\ref{fig:cma} show the location of galaxy models with two
different star formation (SF) histories: a single burst at $z=3$ (dark red)
and an exponentially declining SF starting at $z=3$ (green) with a
characteristic time--scale ($\tau$) of $1$ Gyr.  Both models were calculated
using the population synthesis code by \citet{bc03} with a Chabrier initial
mass function.  For each SF history, three different metallicities are shown:
$0.02$, $0.008$ and $0.004$, from brighter to fainter objects.  The relation
between metallicity and luminosity in these models has been calibrated by
requiring that they reproduce the observed CMR in Coma (see
Sec.~\ref{sec:local}).  This calibration has been found {\it a posteriori} to
be in good agreement with the metallicity--luminosity relation derived from
spectral indices of Coma galaxies \citep{poggianti01b}.  The solid vertical
line in each panel of Fig.~\ref{fig:cma} shows the apparent magnitude which
translates to M$_{\rm V}=-18.2$ when evolved passively to $z=0$ using the
single burst model.  This magnitude limit was chosen so that all brighter
objects are above the $5\sigma$ detection limit in the V--band in all of our
clusters.

Note that the FORS2 field is $6\farcm8\times6\farcm8$ with a pixel size of
$0\farcs20$ and, after dithering, the field of view with the maximum depth of
exposure is approximately $6.5^{\prime} \times 6.5^{\prime}$.  For our infrared
data, however, taking into account dithering and overlapping exposures, the
field is $6.0^{\prime} \times 4.2^{\prime}$ for our intermediate redshift
cluster candidates and $5.4^{\prime} \times 4.2^{\prime}$ for our high redshift
cluster candidates.  As explained in Sec.~\ref{sec:photoz}, galaxies likely to
be cluster members are selected as having a probability within $\pm\,0.1$ of
the cluster redshift above some threshold.  Where there is no IR data, however,
the redshift probability distributions are broader, and so cluster likelihoods
are correspondingly lower and fewer galaxies meet the adopted criteria.  While
we have re--calibrated the photo-z cutoffs in the regions where no IR data are
available so to use broader cuts in these regions, we have noted that some
`edge effects' remain.  For these reasons, in this study we will only use the
region for which we have both optical and infrared data for each of our fields.

The open circles shown in Fig.~\ref{fig:cma} correspond to objects within the
fixed maximum physical radius centred on the BCG and included in the SOFI
field.  This physical radius turns out to be $\sim 0.74\,{\rm Mpc}$.  For two
of the fields shown in Fig.~\ref{fig:cma} (cl$1227.9$-$1138$ and
cl$1103.7$-$1245$a), the BCG lies close to the edge of the chip (see Fig.~1 in
Poggianti et al. 2006 and Fig.~6 in White et al. 2005).  In these cases, the
open circles show all the objects with high probability of cluster membership
within the whole region for which we have infrared data.

\begin{figure}
\bc
\hspace{-0.8cm}
\resizebox{8cm}{!}{\includegraphics{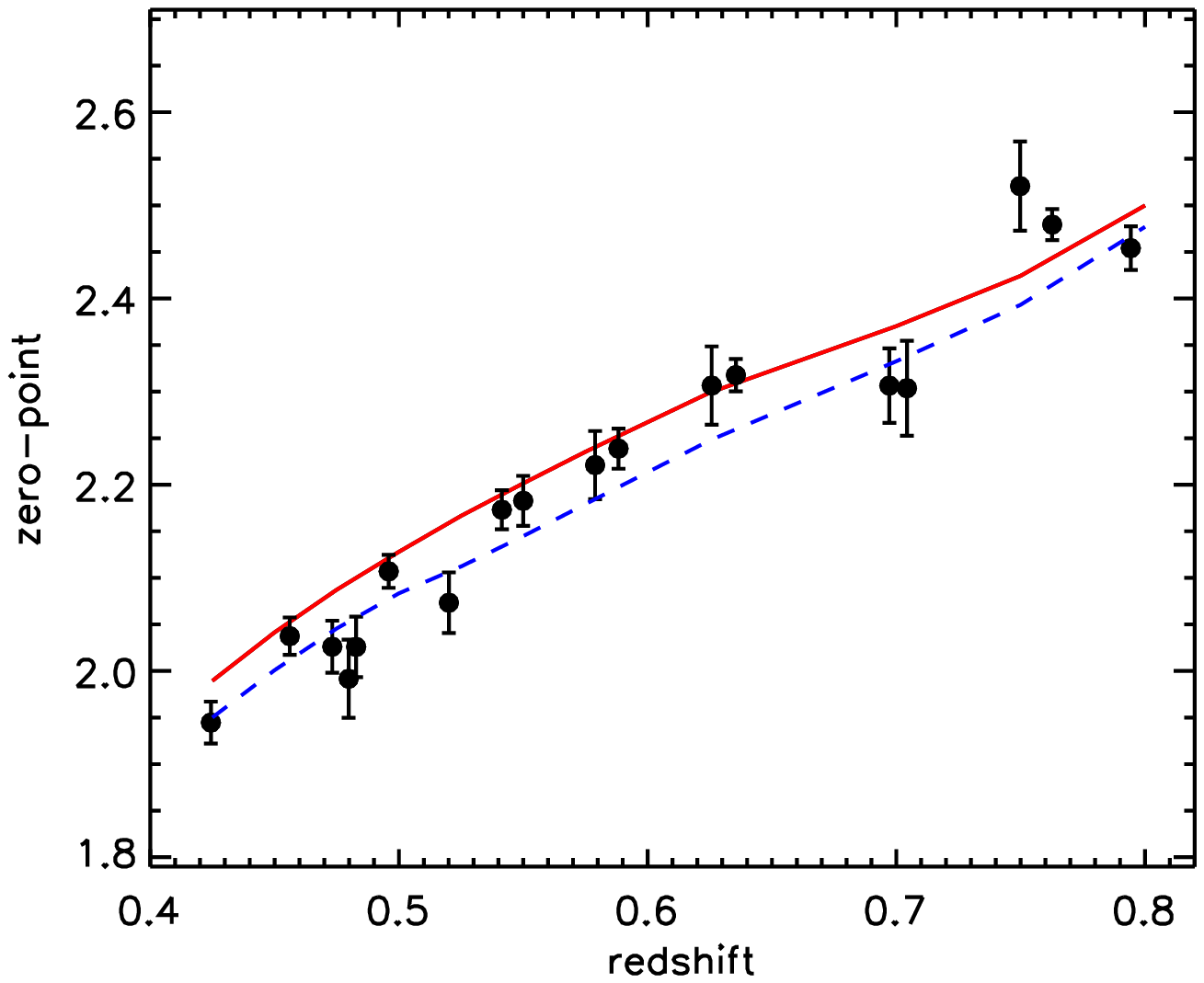}}\\%
\caption{Evolution of the zero-point of the colour--magnitude relation as a
  function of redshift.  Filled circles with error bars represent the values
  measured by fitting the observed relation.  The solid and dashed lines show
  the predictions of a single burst model with formation redshift $3$ and $2$
  respectively.  In both cases, the zero-point has been measured at the
  apparent magnitude that corresponds to $M_{\rm V} = -20.$ when evolved to
  $z=0$.}
\label{fig:zp}
\ec
\end{figure}

Fig.~\ref{fig:cma} shows several interesting results.  The single burst model
seems to provide a good fit to the observed red sequence over all the redshift
range sampled by the clusters under investigation.  This confirms that the
location of the CMR observed in distant clusters, requires high redshifts of
formation, and that the slope is consistent with a correlation between galaxy
metal content and luminosity.  In Fig.~\ref{fig:zp}, we show the evolution of
the zero-point of the colour--magnitude relation as predicted by adopting the
single burst model with formation redshift $3$ (solid line), and the values
measured by fitting the observed relation for all the clusters in our sample
(filled symbols with error bars).  Both for the observational data and for the
model predictions, the zero-point has been measured at the apparent magnitude
that corresponds to $M_{\rm V} = -20$ when evolved to $z=0$.  As mentioned
before, the fit has been obtained using only the spectroscopically confirmed
members without emission lines in their spectra (red filled circles in
Fig.~\ref{fig:cma}) and assuming a fixed slope of $-0.09$.  Error bars have
been estimated via bootstrap resampling of the galaxies used to compute the
fit.  Overall, the measured values follow nicely the model line, although some
deviations are visible where the models are systematically redder than the best
fit relations (see also Fig.\ref{fig:cma}).  These could indicate a lower
formation redshift for red--sequence galaxies in these clusters.  For a single
burst model with formation redshift $2$, the predicted zero-point is lower at
all redshifts, as indicated by the dashed line in Fig.~\ref{fig:zp}.  Within
the errors, however, both single burst models provide a relatively good fit
over all the redshift range under investigation.

Another interesting result shown in Fig.~\ref{fig:cma} was already noted in
\citet{white05}: our sample shows large variations in the relative proportions
of red and blue galaxies.  Some clusters exhibit a strong red sequence with
rather few blue galaxies (e.g.  cl$1232.5$-$1250$, cl$1138.2$-$1133$) while
others show many blue galaxies but relatively few passively evolving systems
(e.g.  cl$1040.7$-$1155$, cl$1227.9$-$1138$).  So although the {\it locus} of
red galaxies is well described by a uniformly old stellar population, the
``morphology'' of the colour--magnitude diagrams is quite varied.  This must be
related, at some level, to the dynamical and accretion histories of the
clusters.

\begin{figure}
\bc
\hspace{-0.8cm}
\resizebox{8cm}{!}{\includegraphics{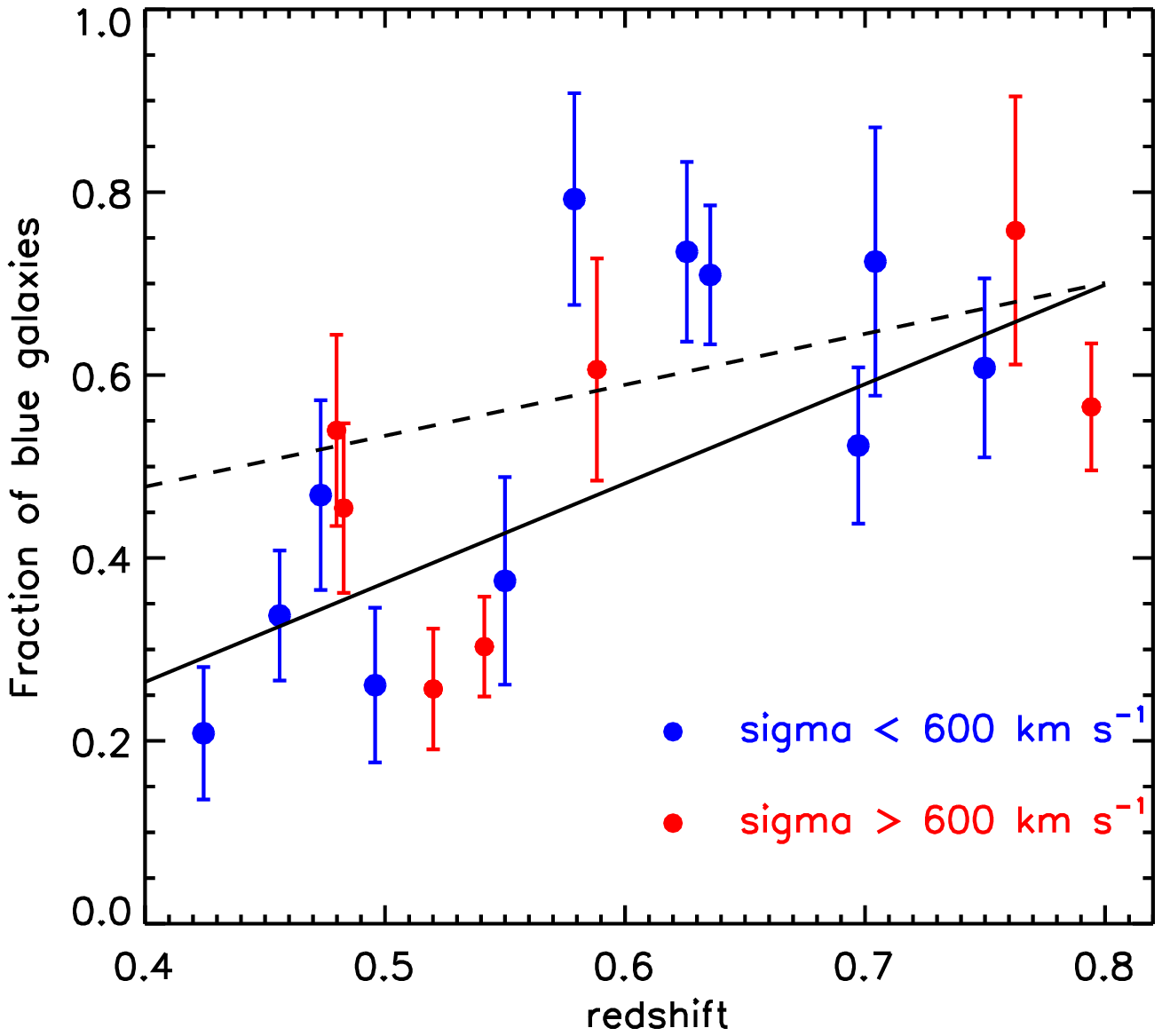}}\\%
\caption{Fraction of galaxies bluer than $0.3$~mag from the best fit colour
  magnitude relation as a function of redshift.  Red and blue symbols are
  for galaxy clusters with velocity dispersion larger and smaller than
  $600\,{\rm km}\,{\rm s}^{-1}$ respectively. Errors have been estimated using
  Poisson statistics. The solid and dashed line show linear fits to the data
  obtained by selecting cluster members using their photometric redshifts and
  statistical subtraction respectively. Points are shown for the former
  selection criterion. (See text for details).}
\label{fig:bf}
\ec
\end{figure}

In Fig.~\ref{fig:bf} we show the fraction of blue galaxies for the $18$ EDisCS
fields shown in Fig.~\ref{fig:cma} as a function of redshift.  Red and blue
symbols are for clusters with velocity dispersion larger and smaller than
$600\,{\rm km}\,{\rm s}^{-1}$ respectively.  The blue fractions shown in
Fig.~\ref{fig:bf} have been computed by counting the galaxies bluer than $0.3$
from the best fit colour--magnitude relation, and brighter than the passive
evolution corrected limit that corresponds to $-18.2$ in the rest-frame V-band
at $z=0$.  All the objects which have a high probability of cluster-membership
and within the same areas used for Fig.~\ref{fig:cma} are used.  The solid line
shows a linear fit to the data, while the dashed line shows a linear fit to
data obtained using a statistical subtraction technique to determine cluster
membership.  Errors have been estimated using Poisson statistics.  Our
definition of blue fraction differs from the original definition introduced by
\citet{bo84}.  It is, however, interesting to note that results shown in
Fig.~\ref{fig:bf} exhibit - as found for the first time by \citet{bo84} - an
increase in the fraction of blue galaxies with increasing redshift.  The trend
is present, although weaker, also when a statistical field subtraction instead
of photometric redshift information is used to determine cluster membership.  A
Spearman's rank correlation test gives a coefficient of $0.50$ with a
significance level of $0.036$ in the case membership is defined using
statistical subtraction, while in the case membership is defined using
photometric redshifts, the coefficient is $0.69$ with a significance level of
$0.001$.  It should be noted, however, that the error bars are quite large and
that there are large cluster-to-cluster variations at any given redshift.

We note that the constraints on the photometric redshift for blue galaxies are
usually worse than those for galaxies of the same magnitude but with a redder
colour, because of their smoother spectral energy distribution.  The
statistical subtraction technique is also more uncertain for blue galaxies
because the field population that is used to perform the subtraction has a
colour distribution peaked towards blue colours.  In the following sections, we
will concentrate on the distribution of galaxies along the red--sequence, where
both the photometric redshift and the statistical subtraction method are
expected to perform better.

%%%%%%%%%%%%%%%%%%%%%%%%%%%%%%%%%%%%%%%%%%%%%%%%%%%%%%%%%%%%%%%%%%%%%%%%%%%%%%%
\section{The red--sequence galaxy distribution}
\label{sec:redshift}

\begin{figure}
\bc
\vspace{-1.cm}
\hspace{0.8cm}
\resizebox{8cm}{!}{\includegraphics{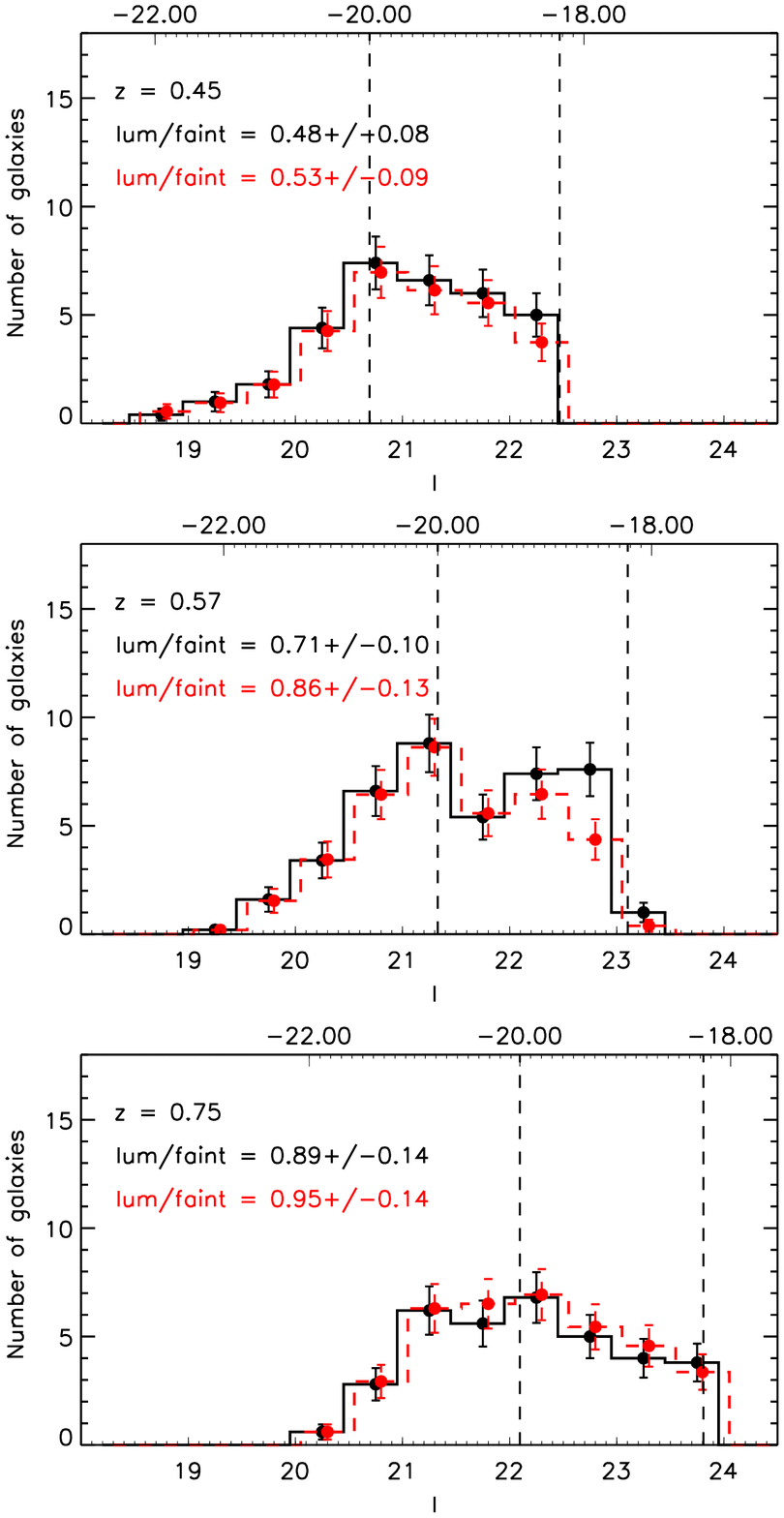}}\\%
\caption{Number of galaxies along the red--sequence.  Histograms have been
  obtained by stacking individual clusters in three redshift bins.  Black and
  red histograms correspond to cluster membership based on photometric redshift
  and on statistical subtraction respectively.  The scale on the top of each
  panel shows the rest--frame V--band magnitude that corresponds to the I--band
  magnitude and has been passively evolved to $z=0$.  Vertical dashed lines
  show our magnitude limit and the edge between `luminous' and `faint'
  galaxies.  A small offset has been added between two histograms in each panel
  for clarity.  For each histogram, we give the luminous--to--faint ratio as
  defined in the text.}
\label{fig:histo}
\ec
\end{figure}

In \citet{delucia04b}, we analysed the distribution of galaxies along the
colour--magnitude relation for four of the highest redshift clusters in the
EDisCS sample.  As mentioned in Sec.~\ref{sec:intro}, our analysis pointed out
a significant {\it deficit} of faint red galaxies compared to the nearby Coma
cluster. We interpreted this deficit as evidence for a large number of galaxies
having moved onto the red--sequence relatively recently, possibly as a
consequence of the suppression of their star formation by the dense cluster
environment.  If the scenario we envisaged is correct, then we should be able
to see some evolution in the relative number of `faint' and `luminous' red
galaxies as a function of redshift within our EDisCS sample.

In Fig.~\ref{fig:histo} we show the number of galaxies along the red--sequence
obtained by averaging the distributions of individual clusters in three
redshift bins.  Black histograms are obtained by selecting all the galaxies for
which our photometric redshift criterion gives a high probability of cluster
membership (Sec.~\ref{sec:photoz}).  Red histograms are obtained by
selecting cluster members using a purely statistical subtraction
(Sec.~\ref{sec:statsub}).  In the latter case, we have averaged over $100$
Monte Carlo realizations of the statistical subtraction for each cluster.  For
the histograms shown in Fig.~\ref{fig:histo}, all the `members' within $\sim
0.5\times{\rm R}_{200}$ and within $\pm\,0.3$~mag of the best fit relation are
used. In the present analysis, we are neglecting the clusters cl$1227.9$-$1138$
and cl$1103.7$-$1245$a for which the BCG lies at the edge of the chip.
Furthermore, we also neglect the cluster cl$1119.3$-$1129$ which, as explained
in \citet{white05}, shows a very weak concentration of galaxies close to the
selected BCG and has a small value of ${\rm R}_{200}$.  As a consequence, there
are just a handful of galaxies on the red sequence within the fraction of
virial radius we have adopted.  In addition, we do not have infrared data for
this cluster.

Clusters have been combined in three redshift bins (we end up with $5$ clusters
in each bin) correcting colours and magnitudes to the central redshift of the
corresponding bin.  Corrections are based on the single burst model shown in
Fig.~\ref{fig:cma} (results do not appreciably change using a single burst
model with formation redshift $2$ instead of $3$).  The scale on the top of
each panel in Fig.~\ref{fig:histo} shows the rest--frame V--band magnitude
corresponding to the observed I--band magnitude after correcting for passive
evolution between the redshift of the bin and $z=0$.

As in \citet{delucia04b}, we compute for each redshift bin a
`luminous--to--faint' ratio.  We classify as `luminous' all galaxies brighter
than $M_{\rm V}= -20.0$ and as `faint' those galaxies that are fainter than
this magnitude and brighter than $M_{\rm V}= -18.2$.  As mentioned in the
previous section, the faint limit has been chosen because it corresponds to the
limiting magnitude in the I--band for which all selected objects are above the
$5\sigma$ detection limit in the V--band.  As for the choice of the magnitude
corresponding to the edge between faint and luminous galaxies, we use $-20$
because, at our highest redshift, it approximately equally divides the range of
cluster galaxy magnitudes covered down to $-18.2$.  Both limits correspond to
values after passive evolution to $z=0$ and are indicated by vertical dashed
lines in Fig.~\ref{fig:histo}.

\begin{figure}
\bc
\hspace{0.8cm}
\resizebox{8cm}{!}{\includegraphics{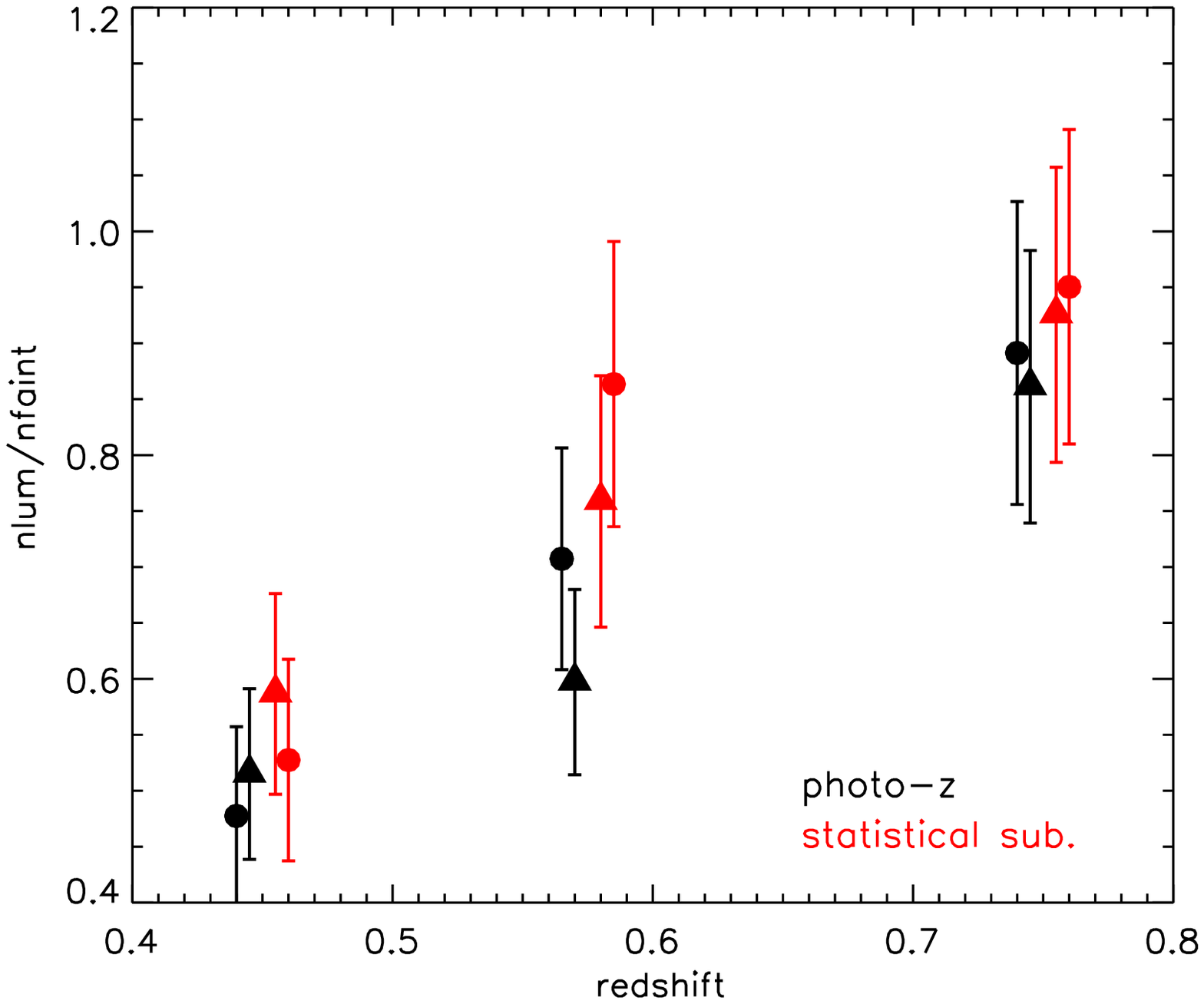}}\\%
\caption{Luminous--to--faint ratio as a function of redshift.  Circles and
  triangles are used in the case where all galaxies within $\sim 0.5\times{\rm
    R}_{200}$ or within $\sim 0.74\,{\rm Mpc}$ from the BCG are selected.
  Black and red symbols correspond to cluster membership based on photometric
  redshift and on statistical subtraction. Symbols corresponding to the same
  redshift have been displaced for clarity.}
\label{fig:ztrend}
\ec
\end{figure}

The values of the luminous--to--faint ratios computed for the histograms shown
in Fig.~\ref{fig:histo} are listed in each panel, along with the estimated
errors.  Fig.~\ref{fig:ztrend} shows these values as a function of redshift.
The error bars have been estimated assuming Poisson statistics and, in the case
where cluster members are selected using a statistical subtraction, they
include the error contribution from the background field (this is however
negligible given the large area used for the subtraction).  Circles in
Fig.~\ref{fig:ztrend} correspond to the histograms plotted in
Fig.~\ref{fig:histo}, where all the cluster members within $\sim 0.5\times{\rm
  R}_{200}$ have been used.  Triangles are used in the case where all the
cluster members within a fixed physical distance ($\sim 0.74\,{\rm Mpc}$) from
the BCG are retained.  Red and black symbols correspond to membership criteria
based on statistical subtraction and on photometric redshifts respectively.

The error bars shown in Fig.~\ref{fig:ztrend} are large, and some small
differences arise from the use of different criteria for cluster membership and
from different choices about the area used for the analysis.  Overall, however,
independently of the method employed and the area used, the data indicate a
decrease of the luminous--to--faint ratio with decreasing redshift.  Faint red
galaxies become increasingly important with decreasing redshift or, in other
words, the faint end of the colour--magnitude relation becomes increasingly
populated with decreasing redshift. As noted in our previous paper, this
finding is inconsistent with a formation scenario in which all red galaxies in
clusters today evolved passively after a synchronous short duration event at
$z\gtrsim 2 - 3$, and suggests that present day passive galaxies follow
different evolutionary paths, depending on their luminosity.

\begin{figure*}
\bc
\vspace{-0.9cm}
\hspace{0.8cm}
\resizebox{18cm}{!}{\includegraphics{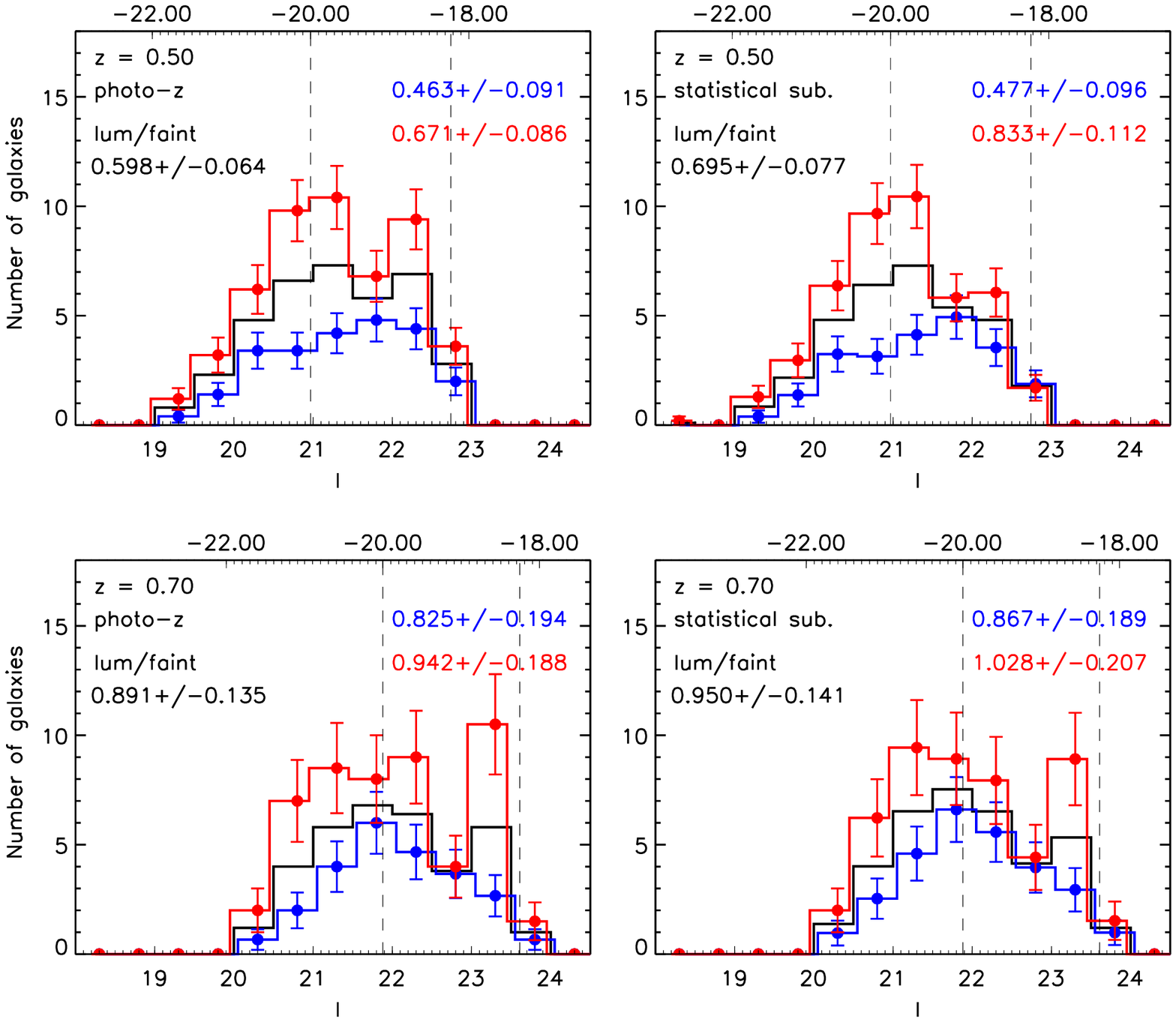}}\\%
\caption{As in Fig.~\ref{fig:histo} for two redshift bins.  Red and blue
  histograms are for clusters with velocity dispersion larger and smaller than
  $600\,{\rm km}\,{\rm s}^{-1}$ respectively.  Black histograms are obtained
  by stacking all the clusters in each redshift bin.  Left and right panels
  correspond to membership based on photometric redshifts and on statistical
  subtraction respectively.  All cluster members within $\sim 0.5\times{\rm
    R}_{200}$ are used. Histograms in each panel have been slightly offset for
  clarity. The luminous--to--faint ratio (with the corresponding errors) are
  listed in each panel.}
\label{fig:sigmatrend}
\ec
\end{figure*}

It is now interesting to ask if this evolution in the luminous--to--faint ratio
depends on cluster properties, for example mass or velocity dispersion.  In
Fig.~\ref{fig:sigmatrend} we again show the distribution of galaxies along the
red--sequence.  This time, we have combined the clusters in two redshift bins
and, in each redshift bin, we have split the clusters according to their
velocity dispersions.  Red and blue histograms are for clusters with
velocity dispersions larger and smaller than $600\,{\rm km}\,{\rm s}^{-1}$
respectively.  Black histograms are obtained by stacking all the clusters in
each redshift bin.  Left panels are for the case where membership is based on
photometric redshifts, while for the histograms shown on the right, membership
is based on a purely statistical subtraction.  All cluster members within $\sim
0.5\times{\rm R}_{200}$ from the BCG are used.  In the lower redshift bin, we
have $5$ clusters in each bin of velocity dispersion while in the higher
redshift bin we have $2$ and $3$ clusters in the larger and smaller velocity
dispersion bin respectively.  The behaviour shown in Fig.~\ref{fig:sigmatrend}
is not significantly different if all galaxies within a fixed physical radius
from the BCG are used (see also Fig.~\ref{fig:ztrend}).

The values listed in Fig.~\ref{fig:sigmatrend} show the same trend of an
increasing luminous--to--faint ratio as a function of redshift and also hint at
a dependence on cluster velocity dispersion.  Clusters with large velocity
dispersion seem to have a larger fraction of luminous galaxies with respect to
the systems with smaller velocity dispersion.  For the highest redshift bin,
the difference goes in the same direction but is not statistically significant.
The number statistics are, however, poor and the error bars are large so that
it is difficult to draw any definitive conclusions.  We find, however, similar
results if we split the clusters on the basis of a richness estimate similar to
that used in \citet{white05}, i.e. based on the number of red--sequence
galaxies.

In a recent study, \citet{tanaka05} have investigated the build up of the
colour--magnitude relation using deep panoramic imaging of two clusters at
$z=0.83$ and $z=0.55$ respectively.  Using photometric redshifts and
statistical subtraction, and using nearest--neighbour density to characterise
the environment, these authors conclude that build-up of the colour--magnitude
relation is ``delayed'' in lower density environments.  This is in apparent
contradiction with our findings, although a direct comparison is not
straightforward as we use the cluster velocity dispersion and not local density
to differentiate environments.  In addition, the conclusions of
\citet{tanaka05} are based only on two clusters and these authors argue that
their intermediate redshift cluster is ``peculiar''.  Finally, one should keep
in mind that cluster--to--cluster variations are rather large (see
Fig.~\ref{fig:cma}).  Further studies are therefore needed to confirm or
disprove the apparent trends.

%%%%%%%%%%%%%%%%%%%%%%%%%%%%%%%%%%%%%%%%%%%%%%%%%%%%%%%%%%%%%%%%%%%%%%%%%%%%%%%
\section{The red--sequence galaxy distribution in nearby clusters}
\label{sec:local}

In the previous section, we have analysed the evolution of the
luminous--to--faint ratio over the redshift range sampled by our EDisCS
clusters.  We want now to set the zero--point for this evolution by studying
the distribution of galaxies along the red--sequence in nearby galaxy clusters.
In order to carry out a comparative study with the clusters in the EDisCS
sample, we need relatively deep photometry (down to $-18.2$ in the
rest--frame V--band) sampling the rest--frame U and V bands, and with good
spatial coverage (at least $\sim 0.5\times{\rm R}_{200}$).  These conditions
are met by only a few nearby clusters.  In the following we will describe in
more detail the data and the analysis performed on our low redshift comparison
sample, which is constituted by Coma, and a sample of clusters selected from
the Sloan Digital Sky Survey (SDSS).

\begin{figure}
\bc
\hspace{0.8cm}
\resizebox{8cm}{!}{\includegraphics{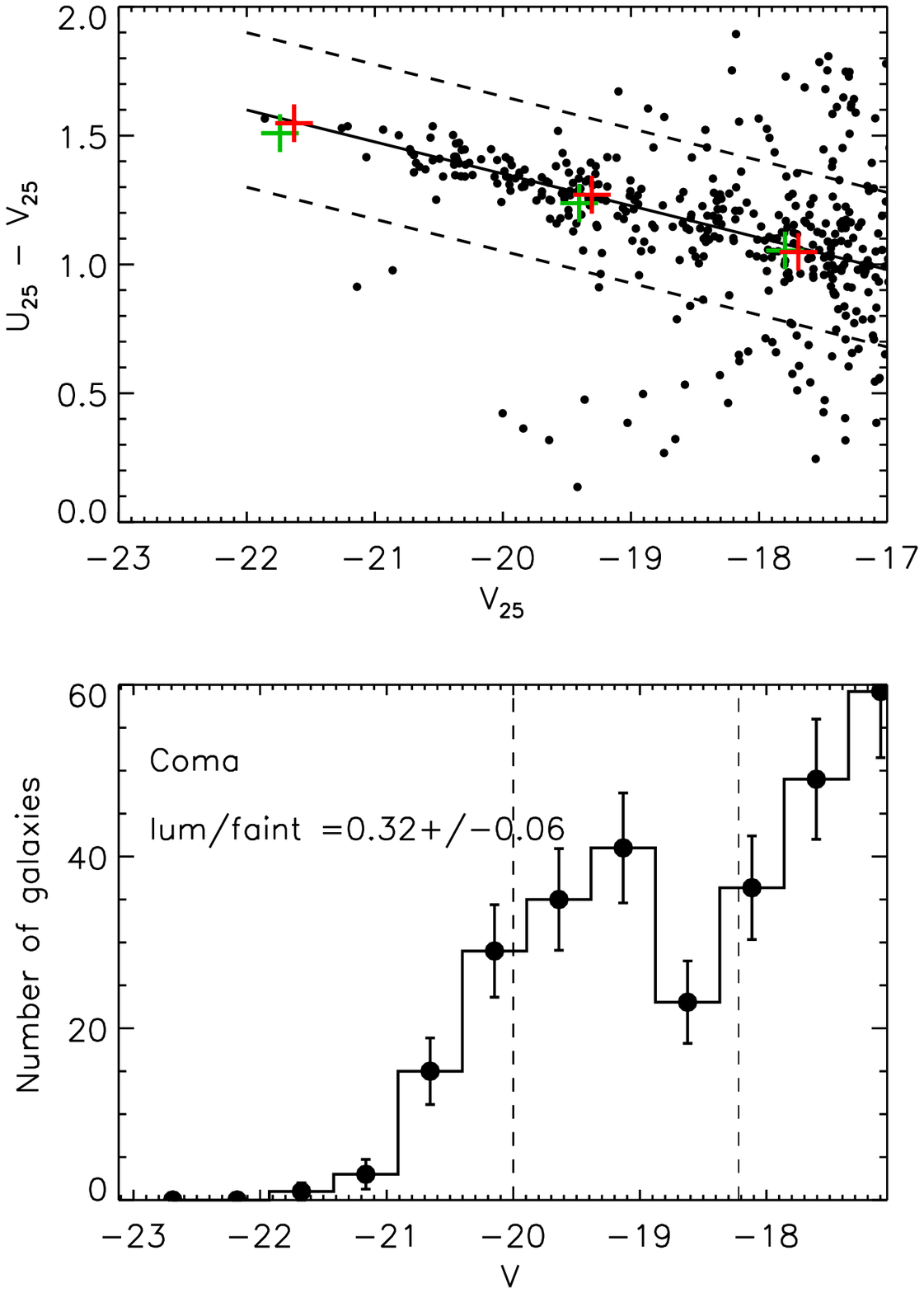}}\\%
\caption{Colour--magnitude relation (top panel) and distribution of galaxies
  along the red--sequence (bottom panel) for the Coma cluster.  The solid line
  in the top panel shows the best fit relation to the red--sequence.  Crosses
  correspond to the same SF models shown in Fig.~\ref{fig:cma}.}
\label{fig:coma}
\ec
\end{figure}

To date, Coma (A1656) remains the only rich cluster in the nearby Universe
($z=0.023$) with a high precision near--UV colour--magnitude relation,
determined using hundreds of spectroscopically confirmed members.  In this
work, we have used magnitudes and colours in a $25\farcs2$ diameter aperture
from \citet*{terlevich01}.  At the redshift of Coma, this corresponds to a
physical size of $11.71$ kpc, quite closely approximating our $\sim 11$--$15$
kpc aperture from $z\sim 0.4$--$0.8$ (see Sec.~\ref{sec:cm}).  Observed
magnitudes are converted to absolute magnitudes using the distance modulus of
Coma computed using its redshift and the adopted cosmology ($35.00$).  Observed
colours are converted to rest--frame colours using tabulated K--corrections
\citep{poggianti97}.  The colour--magnitude relation of the Coma cluster, based
on the catalogue by \citet{terlevich01}, is plotted in the top panel of
Fig.~\ref{fig:coma}.  The solid thick line shows the best fit relation
determined using the bi--weight estimator, as for the EDisCS clusters.  Dashed
lines correspond to $\pm\,0.3$~mag from the best fit line.  Crosses show
predictions from the same two models shown in Fig.~\ref{fig:cma}.  We recall
that the relation between metallicity and luminosity in these models has been
calibrated by requiring that they reproduce the observed CMR in Coma.

The bottom panel of Fig.~\ref{fig:coma} shows the distribution along the
red--sequence for this cluster.  Membership information has been obtained using
a redshift catalogue kindly provided by Matthew Colless and a procedure similar
to that employed in \citet{mobasher03}.  Briefly, for each magnitude bin, we
count how many objects have a measured spectroscopic redshift ($N_z$), and how
many are spectroscopically confirmed members ($N_c$).  We assume then that the
spectroscopic sample is `representative', i.e. that the fraction of galaxies
that are cluster members is the same in the spectroscopic sample (that is
incomplete) as in the photometric sample (that is complete).  Cluster
membership can then be obtained as the ratio between the two numbers computed
before ($N_c/N_z$).  The counts shown in the bottom panel of
Fig.~\ref{fig:coma} are obtained correcting the raw distribution by this
membership factor.  The luminous--to--faint ratio we measure for Coma is
$0.32\,\pm\,0.06$.\footnote{In \citet{delucia04b} we erroneously used a
distance modulus equal to $35.16$ (instead of $35.00$ used here) corresponding
to $H_0 = 65\,{\rm km}\,{\rm s}^{-1}\,{\rm Mpc}^{-1}$.  However, a small bug in
the code we used produced a value of the luminous--to--faint ratio that is not
much off that measured in the present study ($0.34\,\pm\,0.06$).}

\begin{figure}
\bc
\hspace{-0.6cm}
\resizebox{8cm}{!}{\includegraphics{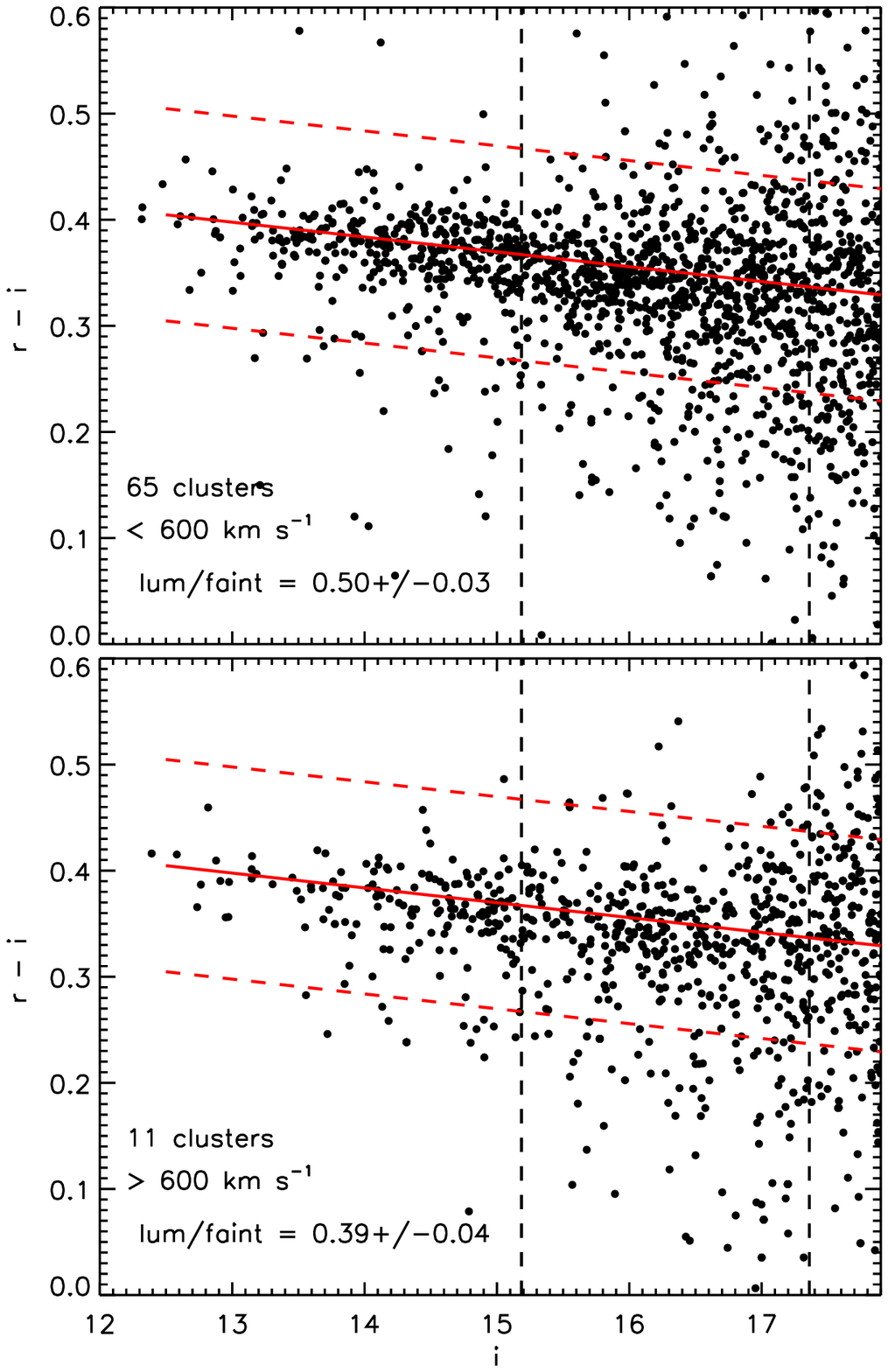}}\\%
\caption{Stacked colour--magnitude relation for galaxy clusters selected from
  the C4 catalogue with velocity dispersion smaller (top panel) and larger than
  $600\,{\rm km}\,{\rm s}^{-1}$.  The solid line in each panel shows the
  colour--magnitude relation predicted by a single burst model with formation
  redshift $3$.  Dashed lines are offset from the solid line by $\pm\,0.1$.
  The vertical dashed lines correspond to ${\rm M}_{\rm V}=-20.18$ and ${\rm
    M}_{\rm V}=-18.25$ respectively. (See text for details).}
\label{fig:sdss}
\ec
\end{figure}

We have complemented our low redshift comparison sample with a sample of
clusters selected from the SDSS. The basis for the cluster sample used here is
the C4 cluster catalogue by \citet{miller05}.  This catalogue is available for
the SDSS Data Release $3$ and is based on the spectroscopic sample.  The
cluster detection algorithm employed for the construction of the C4 catalogue,
is essentially based on an identification in a three-dimensional space
(position, redshift, and colour).  We refer to the original paper for more
details.  Based on the cluster redshifts and velocity dispersions given in the
C4 catalogue, we have re-identified the BCG and measured the velocity
dispersion at the virial radius for each cluster.  Details about the procedure
are described in von der Linden et al. (in preparation).  For the purposes of
this analysis, we have used the $76$ clusters below $z=0.045$ so to assure
completeness down to the magnitude limit used for our analysis.
 
A direct comparison with the analysis of the red--sequence galaxies
distribution performed above for the EDisCS clusters and for Coma is not
simple, and requires a number of steps that we describe in the following.  The
first difficulty comes from the fact that, for the SDSS clusters, we have AB
Petrosian magnitudes, while for the EDisCS clusters and for Coma we have used
aperture magnitudes.  In order to have an estimate of the correction necessary
to convert Petrosian magnitudes into aperture magnitudes, we have compared our
I--band aperture magnitudes to the `total' magnitudes we used in
\citet{white05}.  We recall that an approximate total I-magnitude for each
galaxy was estimated by adding to the Kron magnitude the correction appropriate
for a point source measured within an aperture equal to the galaxy's Kron
aperture (we refer to the original paper for details).  Using the median value
of this correction, and considering that for elliptical like objects Petrosian
magnitudes take into account about $80$ per cent of the light, we estimate that
our aperture magnitudes can be converted into Petrosian magnitudes by a shift
that varies between $-0.05$ and $-0.18$ from fainter to brighter galaxies.

In practice, we have converted the limits used before (${\rm M}_{\rm V}=-18.2$
and ${\rm M}_{\rm V}=-20$) into ``Petrosian limits'' (${\rm M}_{\rm V}=-18.25$
and ${\rm M}_{\rm V}=-20.18$).  The conversion from apparent magnitudes to
absolute magnitudes in the V--band has been performed using the routine {\small
KCORRECT} by \citet{blanton03}.  For each cluster we have then constructed
photometric catalogues by taking all the objects that are classified as
galaxies by the SDSS pipeline and that reside within $\sim 0.5\times {\rm
R}_{200}$ from the BCG.  A corresponding field catalogue has been constructed
by using the whole DR$3$ and for each cluster we have performed $50$ Monte
Carlo realizations of the statistical subtraction procedure described in
Sec.~\ref{sec:statsub}. 

Fig.~\ref{fig:sdss} shows, for one realization of the statistical subtraction,
the stacked colour--magnitude diagrams for the two velocity dispersion bins.
The solid line in each panel shows the relation predicted by a single burst
model with formation redshift $3$, while dashed lines are offset from this line
by $\pm 0.1$~mag.  The vertical dashed lines show the limits used to define the
luminous and faint population.  We note that the magnitudes and colours plotted
in Fig.~\ref{fig:sdss} are given in the AB system.  We have therefore used AB
magnitudes and colours for the single burst model.  The luminous--to--faint
ratio computed from these stacked colour--magnitude relations are listed in
each panel and are $0.39\pm0.04$ for the high velocity dispersion bin, and
$0.50\pm0.03$ for the low velocity dispersion bin.  The value measured for the
clusters with velocity dispersion larger than $600\,{\rm km}\,{\rm s}^{-1}$
appears then compatible, within the errors, with that measured for the Coma
cluster, while the value measured for clusters with lower velocity dispersion
is significantly higher.  We note also that the trend found for the SDSS
clusters is the opposite of what we have found for our high redshift sample.

%%%%%%%%%%%%%%%%%%%%%%%%%%%%%%%%%%%%%%%%%%%%%%%%%%%%%%%%%%%%%%%%%%%%%%%%%%%%%%%
\section{The buildup of the colour--magnitude relation}
\label{sec:build}

\begin{figure}
\bc
\hspace{-0.8cm}
\resizebox{9cm}{!}{\includegraphics{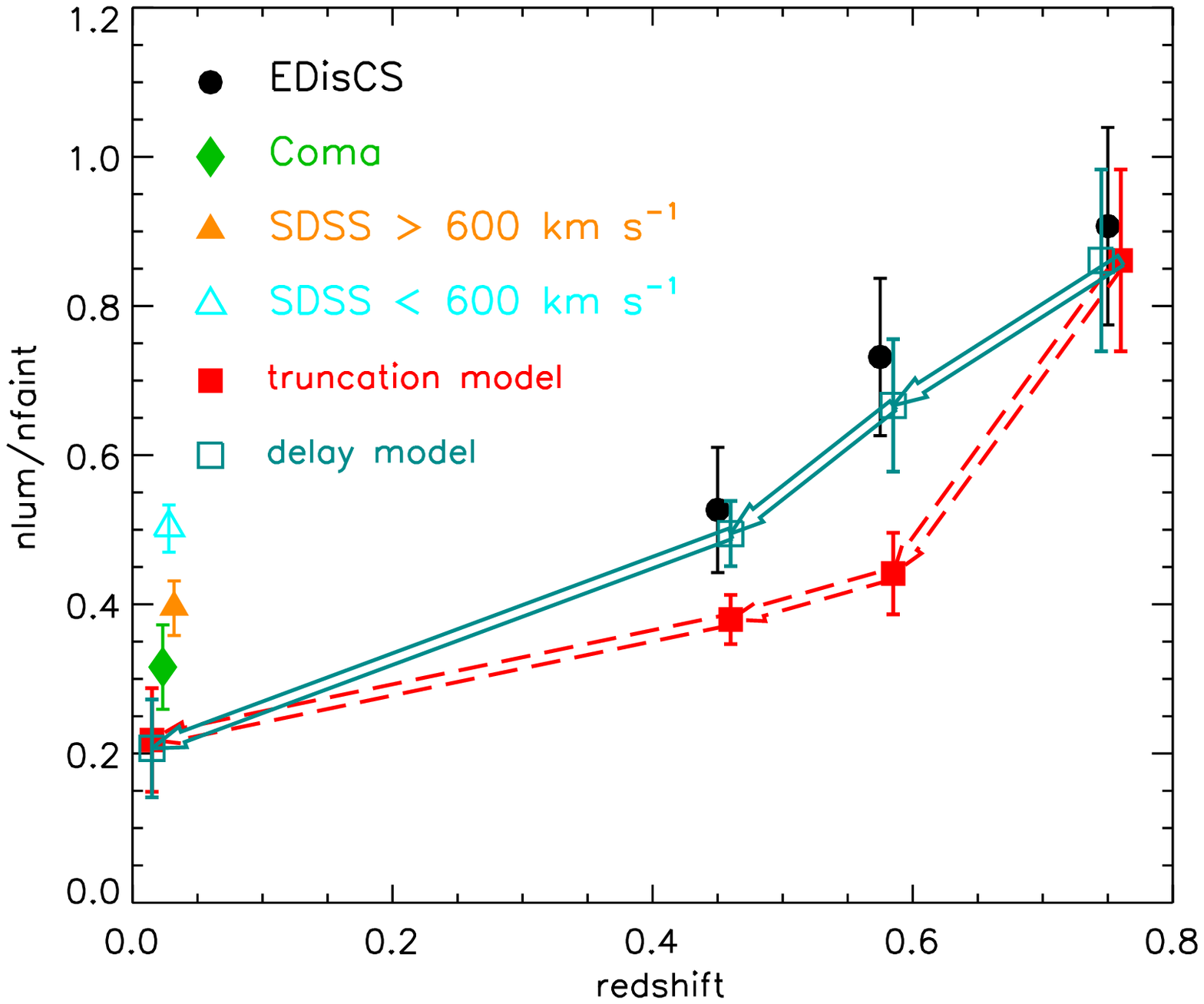}}\\%
\caption{Luminous--to--faint ratio as a function of redshift.  Filled circles
  show the average of the values shown in Fig.~\ref{fig:ztrend}.  The green
  diamond shows the values determined for Coma.  The triangles show results for
  clusters selected from the SDSS with velocity dispersion larger (orange) and
  smaller (cyan) than $600\,{\rm km}\,{\rm s}^{-1}$.  The arrows indicate the
  evolution of the luminous--to--faint ratio obtained using the models
  described in the text.}
\label{fig:zevol}
\ec
\end{figure}

The results of our analysis are summarised in Fig.~\ref{fig:zevol}. Filled
black circles show the values of the luminous--to--faint ratio for EDisCS
clusters in three redshift bins.  The values shown in Fig.~\ref{fig:ztrend},
obtained for different choices of cluster membership criterion and area, have
been averaged together.  The error bars corresponding to these points have also
been averaged, rather than combined in quadrature, in order to give a
`conservative' measure for the uncertainties.  The green diamond shows the
corresponding value for Coma. The orange and cyan triangles show the value
measured for clusters selected from the SDSS with velocity dispersion larger
and smaller than $600\,{\rm km}\,{\rm s}^{-1}$ respectively.

In \citet{delucia04b}, we interpreted the deficit of faint galaxies found in
the high redshift EDisCS clusters, as evidence that a large fraction of
present--day passive faint galaxies must have moved onto the colour--magnitude
relation at redshift lower than $0.8$.  We argued that the population of blue
galaxies observed in distant galaxy clusters provide the logical progenitors of
faint red galaxies at $z=0$.  It is therefore interesting to ask if the
measured evolution in the luminous--to--faint ratio can be reproduced by simple
evolution of the combined blue and red galaxies that populate the
colour--magnitude diagrams of our high redshift clusters.  In order to have a
handle on this question, we have used the population synthesis model by
\citet{bc03} to construct different exponentially declining star formation
histories with $\tau = 1, 2, 3$, and $7$ Gyr and a redshift of formation $3$.
The same metallicities and normalisations used for the models shown in
Fig.~\ref{fig:cma} have been adopted here.  We have then started from the
distribution of galaxies on the colour--magnitude diagram of the clusters in
the highest redshift bin shown in Fig.~\ref{fig:zevol} - for simplicity, we
have used membership based on photometric redshift and a fixed physical
distance from the BCG.  For each galaxy bluer than $0.3$~mag than the best fit
relation, we have determined the ``closest model'' in colour-magnitude space
among those listed above.  Each galaxy is then evolved to the next redshift bin
by the amount predicted by the best fit model with the corresponding star
formation history truncated at the redshift of the observation of the cluster
(truncation model) or $1\,{\rm Gyr}$ later (delayed model).  For the
red--sequence galaxies (those in the stripe used to compute the
luminous--to--faint ratio), we have simply used the single burst model
described in Sec.~\ref{sec:cm}.  Using the same method, we use the distribution
of galaxies in the colour-magnitude diagram of all higher redshift clusters to
predict the evolution to $z=0.45$ and to $z=0.025$.

Model predictions are shown as red filled (truncation model) and open blue
(delayed model) squares connected by arrows in Fig.~\ref{fig:zevol}.
Interestingly, both models predict an amount of evolution from $z = 0.75$ to
$z\sim0$ that is in nice agreement with that measured using the highest
redshift EDisCS clusters and the Coma cluster.  The predicted amount of
evolution between $z\sim 0.4$ and $z\sim 0$ is instead too large to reproduce
the luminous--to--faint ratios measured for the SDSS clusters.  These appear to
be compatible with the luminous--to--faint ratio measured for the EDisCS
clusters at $z\sim 0.4$.  In the truncation model, galaxies move onto the
red--sequence very rapidly so that the predicted luminous--to--faint ratio lies
below the measured value in the intermediate redshift bins.  In the delayed
model, galaxies stay blue longer so that the predicted evolution is closer to
the observed trend at all redshifts sampled by the EDisCS clusters.

\begin{figure}
  \bc \hspace{-0.8cm}
  \resizebox{9cm}{!}{\includegraphics{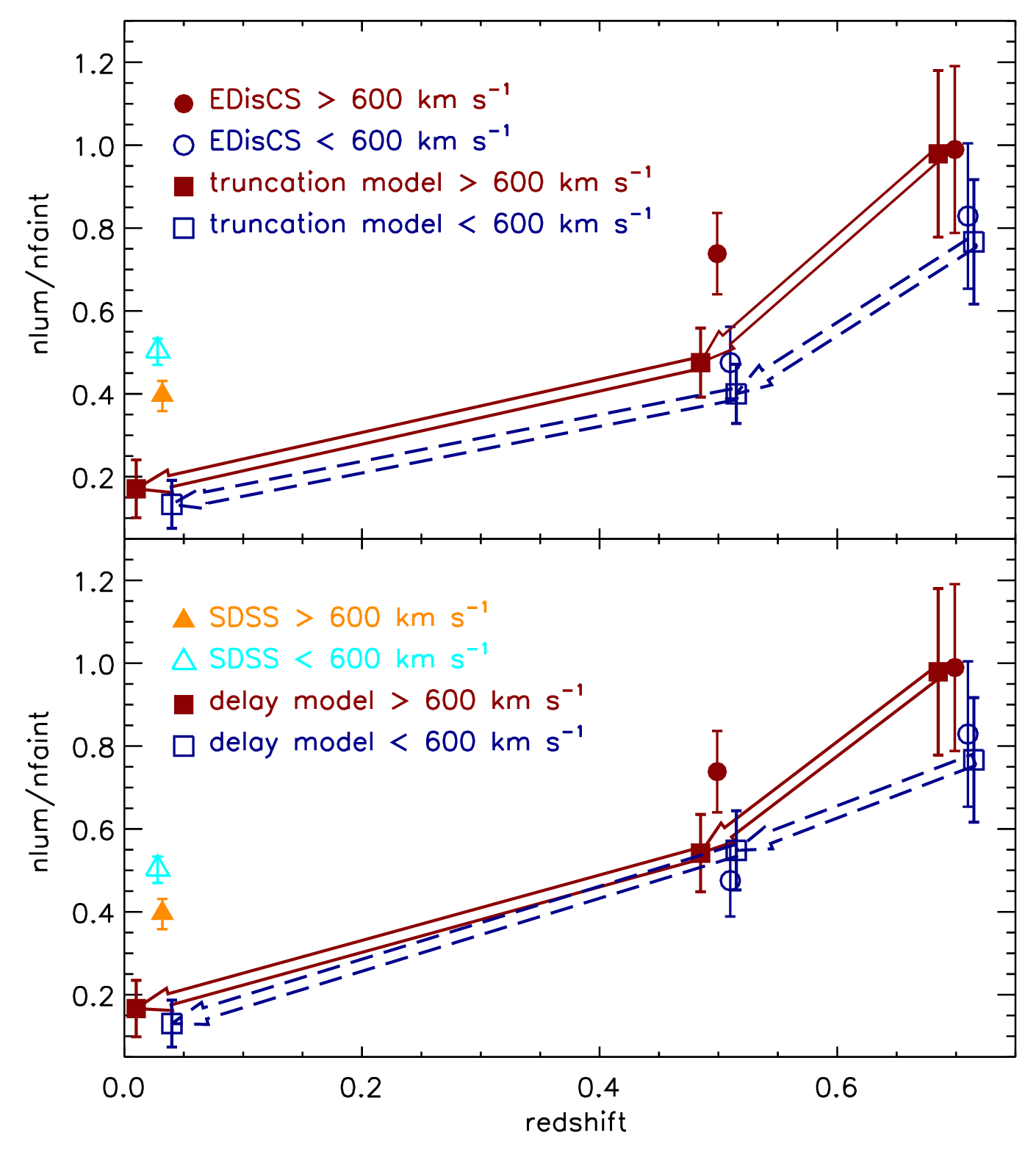}}\\%
\caption{Luminous--to--faint ratio as a function of redshift for clusters
  split according to their velocity dispersions.  Filled red and open blue
  circles are used for EDisCS clusters with velocity dispersions larger and
  smaller than $600\,{\rm km}\,{\rm s}^{-1}$ respectively.  Triangles refer to
  the SDSS clusters as in Fig.~\ref{fig:zevol}.  Squares connected by arrows
  show prediction from the truncation (top panel) and delay (bottom panel)
  described in the text.}
\label{fig:zevolsigma}
\ec
\end{figure}

In Sec.~\ref{sec:redshift} we have investigated how the distribution of
galaxies on the red--sequence depends on the cluster velocity dispersion. The
corresponding luminous--to--faint ratios (again averaged for different cluster
membership criteria and different choices for the area) are shown in both
panels of Fig.~\ref{fig:zevolsigma}.  Filled red and open blue circles are used
for clusters with velocity dispersion larger and smaller than $600\,{\rm
  km}\,{\rm s}^{-1}$ respectively.  The arrows connected by squares show the
evolution predicted by the truncation (top panel) and delay (bottom panel)
models described above.  Triangles refer to the SDSS clusters as in
Fig.~\ref{fig:zevol}.  Both models predict a luminous--to--faint ratio that is
close to observed value for low velocity dispersion clusters at redshift $z\sim
0.5$.  For clusters with velocity dispersion larger than $600\,{\rm km}\,{\rm
  s}^{-1}$, both models instead predict a lower value than that measured for
intermediate redshift EDisCS clusters.  The values predicted for clusters with
larger velocity dispersion are in both models and down to redshift zero, larger
than the corresponding values for low velocity dispersion clusters.  This
appears in agreement with the trend found within our EDisCS, but in
contradiction with that found for the SDSS clusters.  In addition, as shown
already in Fig.~\ref{fig:zevol}, the amount of evolution predicted between
$z\sim 0.5$ and $z\sim 0$ is not compatible with the high luminous--to--faint
ratios measured for the SDSS clusters.

We note that Table~4 in \citet{poggianti06} indicates that a cluster with
velocity dispersion $\sim 600\,{\rm km}\,{\rm s}^{-1}$ at $z\sim 0.6$ evolves
into a system with velocity dispersion $\sim 700\,{\rm km}\,{\rm s}^{-1}$ at
$z\sim 0$.  Therefore, it would be more correct to compare the values predicted
from our models at $z\sim 0$ with those obtained for the SDSS by using a cut
corresponding to $\sim 700\,{\rm km}\,{\rm s}^{-1}$.  In this case, the
measured luminous--to--faint ratios are $0.47\pm0.03$ and $0.44\pm0.06$ for the
low and high velocity dispersion clusters respectively.  The sample used in
this study, however, only contains $4$ cluster with velocity dispersion larger
than $\sim 700\,{\rm km}\,{\rm s}^{-1}$.

%%%%%%%%%%%%%%%%%%%%%%%%%%%%%%%%%%%%%%%%%%%%%%%%%%%%%%%%%%%%%%%%%%%%%%%%%%%%%%%
\section{Discussion and conclusions}
\label{sec:discconcl}

The tight colour--magnitude relation shown by cluster elliptical galaxies has
been the subject of numerous studies that, in the last decade, have been pushed
to higher and higher redshifts.  An interesting and still controversial result
of recent years is the claim of a `deficit' of faint red galaxies at redshift
$z \sim 1$.  In our previous work \citep{delucia04b}, we analysed the
distribution of galaxies on the red--sequence for four high redshift EDisCS
clusters, and we compared such a distribution to that measured for the nearby
Coma cluster.  Although with a low significance level, we found that clusters
at redshift $z \sim 0.8$ exhibit a `deficit' of faint red sequence galaxies.

A decline in the number of red sequence members at faint magnitudes was first
observed in clusters at $z=0.25$ by \citet{smail98}.  Evidence for a
`truncation' of the red--sequence was noted in an overdensity around a
radio galaxy at $z=1.2$ by \citet{kajisawa00} and \citet{nakata01}.  The same
authors, however, speculated that their result might be spurious because
of limited area coverage ($<0.33$ Mpc) and strong luminosity segregation.  (See
also the discussion by \citet{kodama04} who obtain a similar result for
early--type galaxies in a single deep field).   

In recent work, \citet{andreon06} has studied the red--sequence luminosity
function for the cluster MS$1054$--$0321$.  By comparing his result with the
faint--end slope measured by fitting the red--sequence luminosity function of
nearby clusters from the SDSS by \citet{tanaka05}, he concluded that there is
no evidence for a decreasing number of faint red galaxies at higher redshift.
The results obtained by Andreon are based on a single cluster. The fitting
procedure he adopted for the luminosity function of MS$1054$--$0321$ is not the
same as that employed by Tanaka et al., and a comparison based on best fit
parameters is plagued by the well known covariance between errors on ${\rm
  M}_*$ and on $\alpha$.  In addition, we note that Tanaka et al. use an
estimate of local density based on nearest--neighbour statistics.  This
complicates the comparison with MS$1054$--$0321$, a cluster with large velocity
dispersion, bright X--ray luminosity and evident substructures
\citep{donahue98,tran99}.
  
Evolution in the distribution of red--sequence galaxies is expected as a
natural consequence of the recently established mass-dependence of elliptical
galaxy evolution both in clusters and in the field
\citep{thomas04,gallazzi05,vandervel05,treu05,holden05,nelan05}.  The
discussion above suggests the details are still uncertain.  The error bars are
large and large cluster--to--cluster variations preclude definitive
conclusions.  Nevertheless, the build--up of the colour--magnitude relation is
of great interest, as it can constrain the relative importance of star
formation and metallicity in establishing the observed properties of elliptical
galaxies.
  
In this paper, we have used galaxy clusters from our EDisCS sample in order to
extend our previous analysis to a wider redshift range and to study how this
effect depends on cluster velocity dispersion.  In agreement with previous
work, we find that bright red--sequence galaxies in high redshift clusters can
be described as an old, passively--evolving population.  A single burst model
with formation redshift of $3$, calibrated on the colour--magnitude relation of
the nearby Coma cluster, provides a good fit to the red sequence
observed over the full redshift range sampled by our clusters.  This confirms
earlier claims that the location of the red--sequence in distant clusters
suggests high formation redshift, and that its slope is consistent with a
correlation between metal content and luminosity.
  
However, within the same EDisCS sample, we also confirm our previous finding of
a significant evolution in the luminosity distribution of red--sequence
galaxies since $z\sim 0.8$.  Combining clusters in three different redshift
bins, and defining as `faint' all galaxies in the passive evolution corrected
range $0.4\,\gtrsim\,$L$/$L$_*\,\gtrsim\,0.1$, we find a clear decrease in the
luminous--to--faint ratio with decreasing redshift.  The error bars and the
cluster--to--cluster variation are large, but the measured trend is robust
against variations in the criteria adopted for cluster membership and in the
size of the region analysed.  We have also investigated how this evolution
depends on cluster velocity dispersion.  At intermediate redshift, the
luminous--to--faint ratio of clusters with velocity dispersion larger than
$600\, {\rm km}\,{\rm s}^{-1}$ appears to be larger than that measured for
clusters at the same redshift but with lower velocity dispersion.  The error
bars and the cluster--to--cluster variations are, however, too large to draw
any definitive conclusions regarding this point.
  
Our low redshift comparison sample includes the Coma cluster, and a sample of
clusters selected from the SDSS.  For the Coma cluster, we find a value of the
luminous--to--faint ratio that is significantly lower than the value obtained
for the EDisCS clusters at $z\sim 0.45$.  This is not the case for the
luminous--to--faint ratios measured by stacking clusters from the SDSS in
different velocity dispersion bins.  These values are not significantly
different from the values measured for the EDisCS clusters at $z\sim 0.45$.
  
Interestingly, we find the measured amount of evolution in the
luminous--to--faint ratio from $z = 0.75$ to $z\sim0$ to be approximatively
consistent with predictions of simple models where the blue bright galaxies
that populate the colour--magnitude diagram of high redshift clusters, have
their star formation truncated by the hostile cluster environment.  Clearly the
model we use is extremely simplified.  We are assuming a single redshift of
formation for all galaxies and guessing their star formation history simply on
the basis of their location in the observed colour--magnitude diagram.  In
reality, galaxies will have a certain distribution of formation redshifts and
this, together with age, metallicity and dust degeneracies, will certainly
complicate the modelling.  In addition, we are simply assuming that the star
formation history is truncated at the cluster redshift (or 1 Gyr later) for all
galaxies bluer than $0.3$~mag from the best fit red--sequence.  Not all these
galaxies are falling into the cluster at the time of our observations and the
time--scale of the star formation suppression by the cluster environment ({\it
if} there is a suppression of the star formation by the cluster environment)
might be different than that assumed and/or depend on cluster or galaxy
properties.  Finally, we are neglecting further infall of galaxies between our
various redshift bins.  For all these reasons, our model results should be
taken with caution.  They simply suggest that a scenario in which infalling
galaxies have their star formation histories truncated by the hostile cluster
environment is in qualitative agreement with the observed build up of the
colour--magnitude sequence.  They do not yet convincingly confirm this
scenario.
  
Our results indicate that present--day passive galaxies follow different
evolutionary paths, depending on their luminosity (or mass).  This conclusion
is in line with recent results from fundamental--plane and stellar population
studies \citep{thomas04,vandervel05,treu05,holden05}.  More data are required
to clarify if and how this depends on environment.  Such studies will constrain
the relative importance of star formation and metallicity in establishing the
observed red--sequence, and thus clarify the physical mechanisms that drive the
formation and evolution of the early type galaxy population in clusters.

%%%%%%%%%%%%%%%%%%%%%%%%%%%%%%%%%%%%%%%%%%%%%%%%%%%%%%%%%%%%%%%%%%%%%%%%%%%%%%%
\section*{Acknowledgements}
We thank H. McCracken and M. Colless for providing us with electronic
catalogues, and the referee, Vincent Eke, for useful comments. GDL acknowledges
financial support from the Alexander von Humboldt Foundation, the Federal
Ministry of Education and Research, and the Programme for Investment in the
Future (ZIP) of the German Government and the hospitality of the Osservatorio
Astronomico di Padova, where part of this work has been completed.

\bsp

\label{lastpage}

\bibliographystyle{mn2e}
\bibliography{cm_delucia}

\end{document}